\documentclass[preprint2]{aastex}

\begin{document}

\title{Detection of the Cosmic Far-Infrared Background \\
in the AKARI Deep Field South}

\author{S. Matsuura\altaffilmark{1}, M. Shirahata\altaffilmark{1}, M. Kawada\altaffilmark{2}, 
T.T. Takeuchi\altaffilmark{3}, D. Burgarella\altaffilmark{4}, D.L. Clements\altaffilmark{5}, 
W.-S. Jeong\altaffilmark{6}, H. Hanami\altaffilmark{7}, S.A. Khan\altaffilmark{8}, 
H. Matsuhara\altaffilmark{1}, T. Nakagawa\altaffilmark{1}, S. Oyabu\altaffilmark{1}, 
C.P. Pearson\altaffilmark{9}, A. Pollo\altaffilmark{10}, S. Serjeant\altaffilmark{11}, 
T. Takagi\altaffilmark{1}, G. White\altaffilmark{11} }

\email{matsuura@ir.isas.jaxa.jp}

\altaffiltext{1}{Institute of Space and Astronautical Science, JAXA, 3-1-1 Yoshinodai, 
Sagamihara, Kanagawa 229-8510, Japan}
\altaffiltext{2}{Graduate School of Science, Nagoya University, Furo-cho, Chikusa-ku, 
Nagoya 464-0021, Japan}
\altaffiltext{3}{Institute for Advanced Research, Nagoya University, Furo-cho, Chikusa-ku, 
Nagoya 464-8601, Japan}
\altaffiltext{4}{Laboratoire d'Astrophysique de Marseille, Observatoire Astronomique 
Marseille Provence, Universit«e d'Aix-Marseille, CNRS}
\altaffiltext{5}{Physics Department, Imperial College London, Prince Consort Road, 
London SW7 2AZ, UK}
\altaffiltext{6}{Space Science Division, Korea Astronomy and Space Science Institute, 61-1, 
Whaam-dong, Yuseong-gu, Deajeon, 305-348, South Korea}
\altaffiltext{7}{Physics section, Faculty of Humanities and Social Sciences, 
Iwate University, Japan}
\altaffiltext{8}{Shanghai Key Lab for Astrophysics, Shanghai Normal University, 
Shanghai 200234, China}
\altaffiltext{9}{Space Science and Technology Department, Rutherford Appleton Laboratory, 
Chilton, Didcot, Oxfordshire OX11 0QX, UK}
\altaffiltext{10}{The Andrzej Soltan Institute for Nuclear Studies, ul. Hoza 69, 
00-681 Warszawa, Poland}
\altaffiltext{11}{Department of Physics and Astronomy, Open University, Walton Hall, 
Milton Keynes MK7 6AA, UK}

\begin{abstract}
We report the detection and measurement of the absolute brightness and spatial 
fluctuations of the cosmic infrared background (CIB) with the AKARI satellite. 
We have carried out observations at 65, 90, 140 and 160 $\micron$ as a cosmological 
survey in AKARI Deep Field South (ADF-S), which is one of the lowest cirrus regions 
with contiguous area on the sky. After removing bright galaxies and subtracting 
zodiacal and Galactic foregrounds from the measured sky brightness, we have 
successfully measured the CIB brightness and its fluctuations across a wide 
range of angular scales from arcminutes to degrees. The measured CIB brightness 
is consistent with previous results reported from COBE data but significantly higher 
than the lower limits at 70 and 160 $\micron$ obtained with the Spitzer satellite from 
the stacking analysis of 24-$\micron$ selected sources. The discrepancy with 
the Spitzer result is possibly due to a new galaxy population at high redshift 
obscured by hot dust. From power spectrum analysis at 90 $\micron$, three 
components are identified: shot noise due to individual galaxies; Galactic cirrus 
emission dominating at the largest angular scales of a few degrees; and 
an additional component at an intermediate angular scale of $10-30$ arcminutes, 
possibly due to galaxy clustering. The spectral shape of the clustering component 
at 90 $\micron$ is very similar to that at longer wavelengths as observed by Spitzer 
and BLAST. Moreover, the color of the fluctuations indicates that the clustering 
component is as red as Ultra-luminous infrared galaxies (ULIRGs) at high redshift, 
These galaxies are not likely to be the majority of the CIB emission at 90 $\micron$, 
but responsible for the clustering component. Our results provide new constraints 
on the evolution and clustering properties of distant infrared galaxies.
\end{abstract}


\section{Introduction}
Since the Cosmic Infrared Background (CIB) was observed by the COBE 
satellite, it has been known that a large fraction of radiation energy in the Universe 
is released in the infrared \citep{puget96,hauser98}. A compelling explanation for the
CIB in the far-infrared is the thermal emission of interstellar dust associated with 
high-redshift galaxies, heated by the internal UV and optical radiation from 
young stars and active galactic nuclei (AGN), while the near-infrared background is 
dominated by starlight from extragalactic sources \citep{hauser01,kashlinsky05,
lagache05}. Thus, the far-infrared CIB is a powerful probe of dust enshrouded 
star-formation and AGN activity. Measuring the CIB in the far-infrared may constrain 
dust emission powered by pre-galactic objects at very high redshift \citep{cooray04}, 
such as the first generation of stars \citep{bond86,santos02}, which are expected 
to contribute to the near-infrared excess of CIB \citep{matsumoto05}. 

\begin{figure}[!ht]
\begin{center}
   \resizebox{1.0\hsize}{!}{
     \includegraphics*[0,0][900,600]{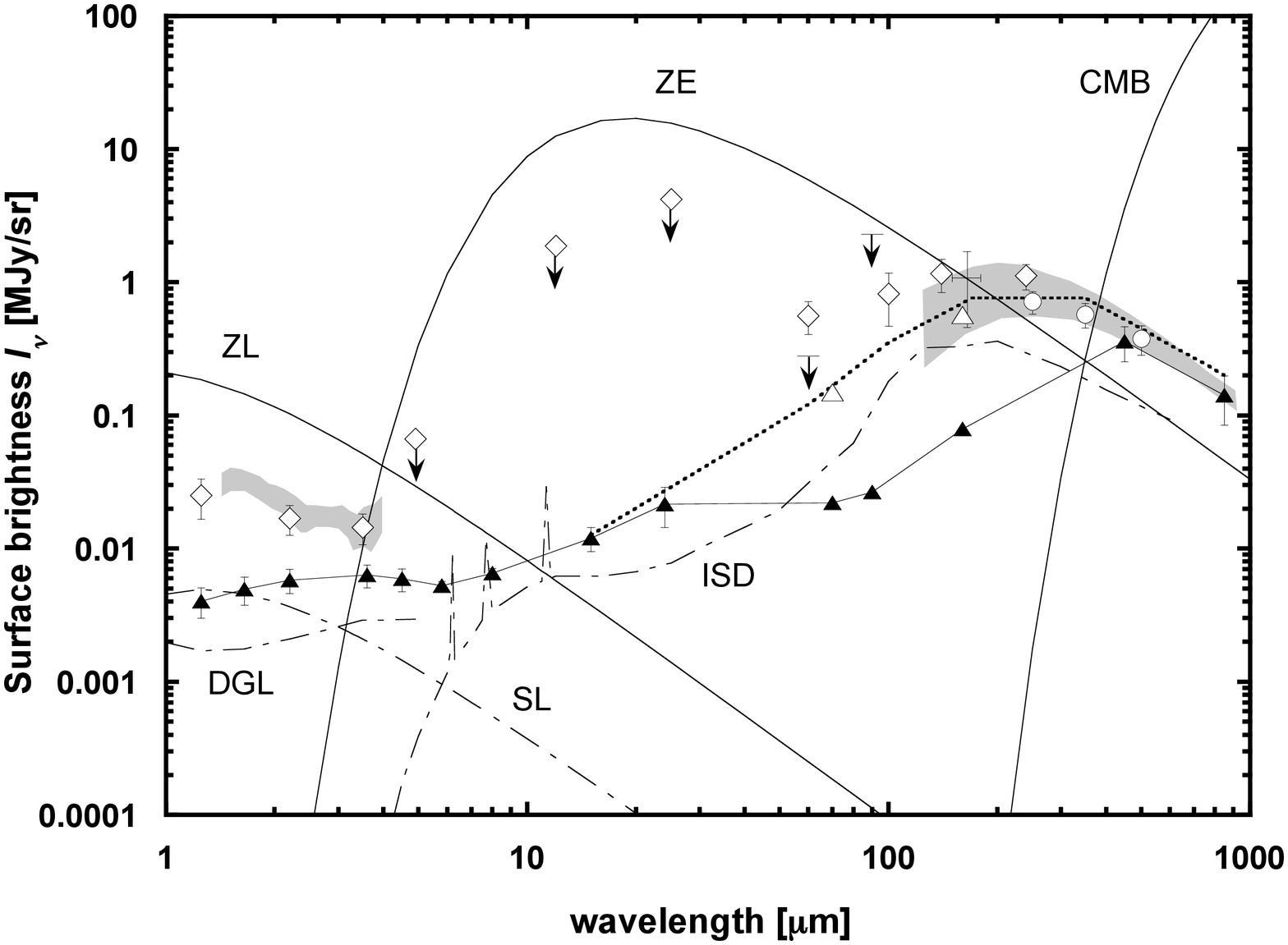}
   }
\caption{
Summary of the results of previous CIB measurements in the infrared and 
submillimeter wavelength range with space missions; COBE/DIRBE (diamonds) 
\citep{hauser98,finkbeiner00}, COBE/FIRAS (shaded region in the submillimeter) 
\citep{fixsen98}, IRAS and ISO (thin line with downward arrows at 60 and 
90 $\micron$ and cross in 150-180 $\micron$) \citep{juvela09}, and IRTS 
(shaded region in the near-infrared) \citep{matsumoto05}. 
For comparison, we also show various foreground emission components at 
dark sky; zodiacal light (ZL), zodiacal emission (ZE), Starlight (SL, $K>9$ mag), 
Diffuse Galactic light (DGL) and Galactic cirrus (ISD), and CMB.
The integrated flux from the galaxy counts by deep surveys from the ground 
in the near-infrared and submillimeter and with the space telescopes in 
the mid- and far-infrared; ISO, Spitzer and AKARI  \citep{kawara98,puget99,
matsuhara00,kawara04,dole04,frayer06a,matsuura07,shirahata08}, are indicated 
by the filled triangles connected with thin lines. 
Stacking of the 24 $\micron$ galaxies for the Spitzer/MIPS map (open triangles) 
and the BLAST map (open circles) results in good agreement with the predicted 
CIB level from a galaxy evolution model by \citet{lagache04} (dotted line). 
}\label{fig:cib_intro}
\end{center}
\end{figure}

In Figure~\ref{fig:cib_intro} we summarize the results of the CIB measurements to date
over the entire infrared and submillimeter wavelength range from previous space 
missions; COBE, IRTS, ISO \citep{hauser98,fixsen98,finkbeiner00,matsumoto05,
juvela09}. For comparison, we also show various foreground emission components; 
zodiacal light (sunlight scattered by interplanetary dust), zodiacal emission 
(thermal emission from interplanetary dust), Galactic light (unresolved stars), 
diffuse Galactic light (starlight scattered by interstellar dust) and Galactic 
cirrus emission (thermal emission from interstellar dust), and the CMB. 
Note at mid-infrared wavelengths it is currently impossible to detect the CIB 
because the zodiacal emission is too bright. In the near-infrared and far-infrared, 
the foreground emission is relatively weak, and careful modeling and subtraction 
of the foreground enables one to extract the CIB from the measured sky brightness.

As seen in Figure~\ref{fig:cib_intro}, the CIB spectrum at wavelengths 
longer than 200 $\micron$ has been well constrained with the FIRAS instrument on COBE 
\citep{puget96,fixsen98}. However, results of the photometric measurements at wavelengths 
shorter than 200 $\micron$ with the DIRBE instrument on COBE  are divergengent
in the mean levels of the CIB brightness, mainly due to the strong and uncertain 
foreground contamination of zodiacal emission, which dominates the sky 
brightness over all of the sky even though the Galactic foreground may be sufficiently 
weak in low cirrus regions \citep{hauser98,lagache00a,finkbeiner00}. 
Although the CIB brightness was recently estimated using the ISOPHOT data, 
independently from the COBE data, the 90 $\micron$ data gave only an upper limit, 
and the measurement accuracy of the CIB brightness in the 150-180 $\micron$ range 
was in fact worse than that of COBE \citep{juvela09}. Figure~\ref{fig:cib_intro} 
clearly shows a wavelength desert of the CIB measurement at shorter 
 far-infrared wavelengths, i.e., $<$200 $\micron$. Hence, new measurements
of the mean level of CIB are demanded in this region.

In the last decade, many observational efforts have been made to resolve the CIB 
into individual galaxies by far-infrared deep surveys with infrared space telescopes 
such as ISO, Spitzer and AKARI \citep{kawara98,puget99,matsuhara00,kawara04,
dole04,frayer06a,matsuura07,shirahata08}, and consequently the origin of CIB has 
become clear. As shown in Figure~\ref{fig:cib_intro}, however, the detected galaxies 
account for only a small fraction ($\sim10\%$) of the measured CIB brightness in 
the far-infrared. \citet{frayer06b} claimed that they resolved more than half of 
the model CIB at 70$\micron$ into point sources in a single deep survey towards 
the GOODS-N field. In the mid-infrared, a lower limit of the CIB at 24 $\micron$ 
is derived from the integrated number counts, and this is thought to account for 
$\sim70\%$ of the model CIB at 24 $\micron$ \citep{papovich04}. \citet{dole06} 
obtained lower limits for the CIB at 70 and 160 $\micron$ by stacking analysis of 
24-$\micron$ detected sources, finding that the mid-infrared sources contribute 
$\sim80\%$ of the CIB in the far-infrared, shown in Figure~\ref{fig:cib_intro} by 
the dotted line. In the submillimeter range, similar stacking analysis of 24-$\micron$ 
galaxies against the deep surveys at 250, 350 and 500 $\micron$ by 
the Balloon-borne Large-Aperture Submillimeter Telescope (BLAST) experiment 
resulted in the conclusion that the 24-$\micron$ sources produce almost all the CIB 
in the submillimeter measured with FIRAS \citep{devlin09, marsden09}. Although 
these studies using the Spitzer  24-$\micron$ surveys provided strong constraints 
on the mean CIB level, the current limit of direct measurement of CIB as diffuse 
emission in the far-infrared range is still high enough to allow the existence of 
new population.

Measuring the spatial fluctuations (anisotropy) of the CIB is a powerful method to 
investigate the unresolved galaxy population below the detection limit with little 
contamination from the foreground. The angular power spectrum of the CIB 
fluctuations is an important observable to trace the distribution of star-forming 
galaxies with respect to the clustering bias in dark matter halos. The fluctuation 
measurement is especially effective at longer wavelengths, where direct 
measurement of the clustering of resolved galaxies is hampered by galaxy 
confusion due to the diffraction-limited resolution of current telescopes. In fact, power 
spectrum analysis of the CIB fluctuations has already been done for large area 
surveys in low cirrus regions; the FIRBACK survey at 170 $\micron$ by ISOPHOT 
\citep{puget99,lagache00b}, the MIPS/Spitzer data at 160 $\micron$ in the SWIRE/GTO 
field \citep{lagache07}, and the BLAST experiment \citep{viero09}. These results provided 
us with useful constraints on galaxy evolution and clustering. Information on the
spectral energy distribution (SED) of the CIB fluctuations would be a further important 
clue in the investigation of the CIB properties. However, the quality of previous 
CIB fluctuation measurements at shorter far-infrared wavelengths, such as 
the ELAIS/Lockman survey at 90 $\micron$ by ISOPHOT \citep{matsuhara00} 
and new analysis of the IRAS map at 60 $\micron$ \citep{miville07}, is not sufficient 
 to add new information over the MIPS and ISOPHOT results at 160 
and 170 $\micron$. 

In this work we aim to make the best measurements of the absolute brightness and 
fluctuations of the CIB since COBE, especially in the shorter far-infrared wavelength range 
which is as yet not extensively explored. This measurement is achieved using the high 
sensitivity of AKARI for diffuse far-infrared emission, high angular resolution, and 
intensive selection of the survey field that provides the minimum foreground contamination. 
In the next two sections we describe the observational method and the data 
processing to produce the final calibrated image of the diffuse emission. In Section 4, 
we describe how the foreground subtraction so that the absolute brightness of CIB 
remains as an isotropic residual. In Section 5, we describe the power spectrum 
analysis of the final image used to decompose the CIB fluctuations into galactic and 
extragalactic components. In Section 6, we show the results of the data analysis 
compared with previous studies of the CIB with scientific discussion.

\section{Observations}
\subsection{Field selection}
The AKARI observations of the CIB were carried out for a far-infrared 
cosmological survey program (PI: S. Matsuura) as one of core scientific programs 
of the AKARI project, known as Mission Programs (MP) \citep{matsuhara06}. 
The survey field named the AKARI Deep Field South (ADF-S) is one of the lowest 
cirrus regions with contiguous area on the sky  \citep{schlegel98} with an area of 
12 square degrees near the South Ecliptic Pole; the field center is located at 
RA=$4^{h}44^{m}00^{s}$, DEC=$-53\degr20\arcmin00\arcsec$ (J2000). 
The survey field includes a nearby cluster of galaxies, DC 0428-53/Abell S0463 
\citep{dressler80}. The observation log of this survey is summarized in Table~\ref{tbl-1}.

The ADF-S field is one of the the best cosmological windows because of its low foreground 
emission; not only a minimum of galactic foreground, but also low zodiacal 
background. High visibility of this field for a polar-orbiting satellite such as AKARI is 
also important to achieve high sensitivity from long exposures. By comparison, the
Lockman hole, the most popular low cirrus region as a cosmological window, 
has a relatively high zodiacal brightness and low visibility due to its low ecliptic latitude.

\subsection{AOT and parameters}
In order to make a mosaic map of the entire field, a total of 200 slow-scan 
observations were obtained by combination of the Astronomical Observation 
Templates (AOTs), FIS--01 and FIS--02 \citep{kawada07,matsuura07}. These AOTs 
provide for multiple pointings as footprints of the field-of-view (FOV) along a scan 
path. The area covered by single observation of FIS--01 and FIS--02 is 
12 arcmin$\times$0.6 deg and 8 arcmin$\times$1.25 deg, respectively, 
for an exposure time of 10 minutes. Data in four photometric bands, centered 
at 65, 90, 140 and 160 $\micron$, were simultaneously taken by a single observation, 
and the same sky is observed in all the bands except for a small difference in the FOV. 

The AOTs include the calibration sequence, i.e. dark current measurement by 
closing the cold shutter and responsivity check with an internal calibrator, 
during maneuvers before and after the observation and at the turning points of 
the round-trip scans. Such a highly redundant calibration sequence enabled us 
to correct any responsivity change, referring to astronomical calibration data taken 
by the same AOT. After the first maneuver to the target position, the cold shutter was 
opened, the sky signals are monitored during a settling time for the attitude control 
while the telescope's line of sight is oscillating around the target position, and then 
the slow-scan observation starts. The signal monitoring before the observation 
is useful to check for any signal drift including time variation of any stray light 
(which is described in the next section).

In order to produce a mosaiced image of the entire field with long-strip image 
segments, a total of 200 slow-scan observations with field overlaps of half a FOV 
in the cross-scan direction were carried out. The survey was completed in one and 
a half years over three observational seasons as summarized in Table~\ref{tbl-1}. 
Mosaicking of the data taken in different seasons is not straightfoward, and 
required careful treatment for the time variation of the zodiacal brightness.

\subsection{Supplemental observations}
In addition to the ADF-S survey,  supplementary observations using 
AOT FIS--01 were also made over small areas towards the Lockman Hole (LH) and 
the North Ecliptic Pole (NEP), as shown in Table~\ref{tbl-1}. One of the LH observations 
was a mini survey to map over $\sim$1 sq.deg for the performance verification (PV) 
soon after the launch of AKARI \citep{matsuura07}, and another was included in 
the same MP program as the ADF-S  to measure the galaxy confusion and to check 
the reproducibility of the measurement a year later. The NEP observation was 
a Director's Time program to evaluate the confusion noise in a relatively higher 
cirrus region compared to the ADF-S, the lowest cirrus region.  As described in the
latter sections, these observations are useful to check the validity of the foreground 
estimate and to evaluate the cirrus contribution to the sky brightness.

\begin{figure}[!ht]
\begin{center}
   \resizebox{1.0\hsize}{!}{
     \includegraphics*[0,0][900,600]{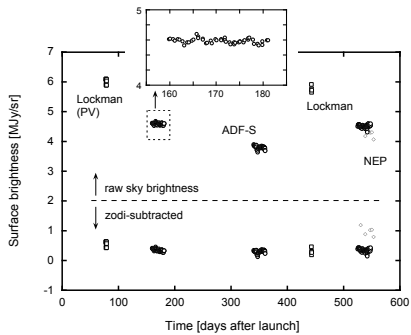}
   }
\caption{
Sky brightness at 65 $\micron$ in ADF-S, Lockman hole and NEP fields, 
measured in various seasons (upper data points) and the residual 
brightness after subtraction of the zodiacal foreground (lower data points). 
The sky brightness values are only for those dates when the fields were 
observed, rather than the whole length of the mission. The inset shows 
a magnified view of the data boxed with the dotted line.
}\label{fig:time_var65}
\end{center}
\end{figure}

\begin{figure}[!ht]
\begin{center}
   \resizebox{1.0\hsize}{!}{
     \includegraphics*[0,0][900,600]{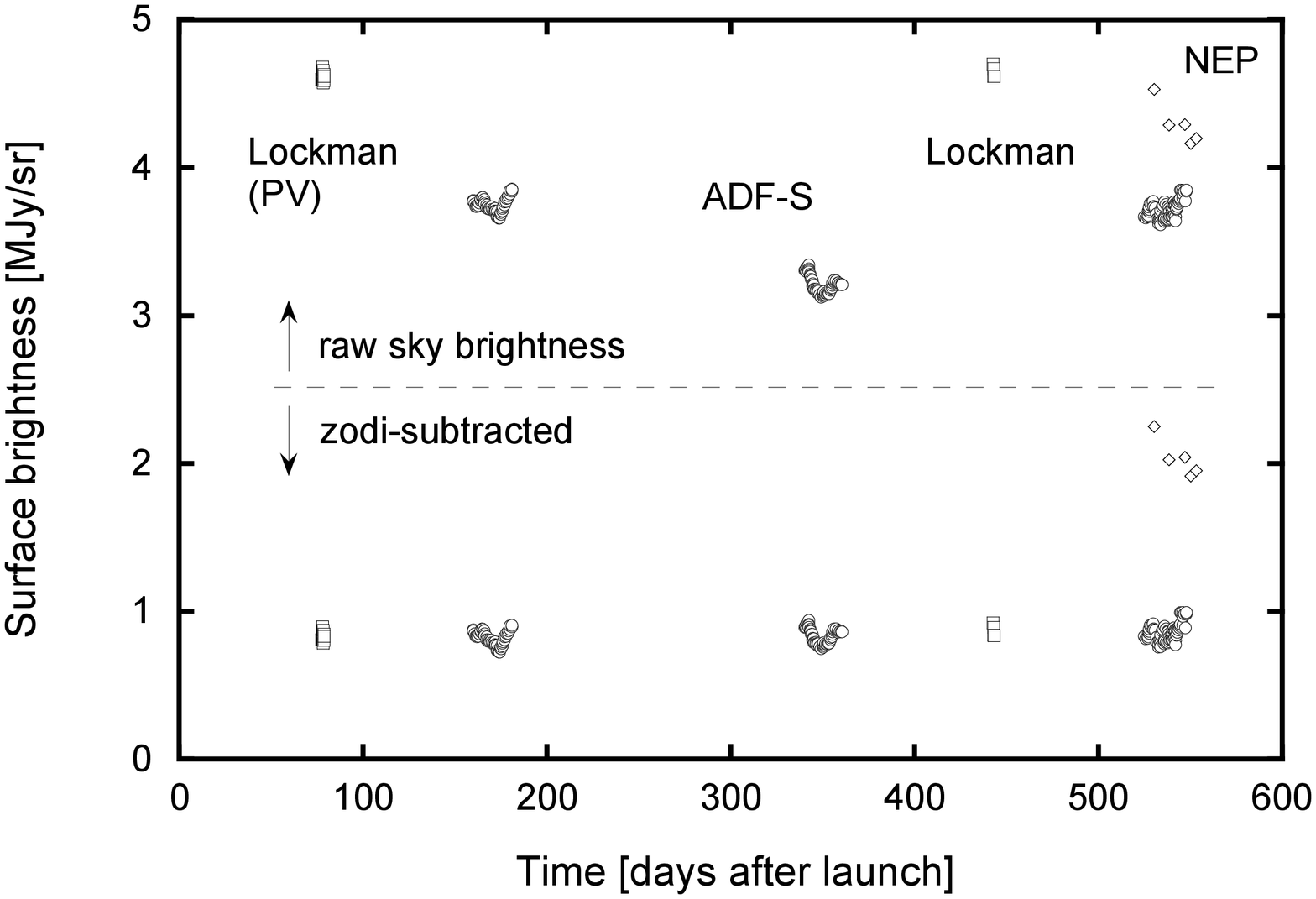}
   }
\caption{
Sky brightness at 90 $\micron$ measured in various fields same as 
Figure~\ref{fig:time_var65}. 
}\label{fig:time_var90}
\end{center}
\end{figure}

\section{Data processing and calibration}
In this section, we describe the methods of data reduction and analysis of the ADF-S data. 
The time series data are processed into a co-added image, using the FIS slow-scan 
analysis tool (SS-Tool, Ver.20070914) \citep{afdum}. The SS-Tool includes dark current 
subtraction, absolute calibration, flat fielding, stray light removal, and co-addition 
of multiple scans. The  SS-Tool is described in detail in \citet{matsuura07}, 
therefore we describe them only briefly in the following sections.

\subsection{Time domain analysis} \label{sec: time domain analysis}
The raw output of the FIS instrument is time series data, consisting of 
integration ramps for each detector pixel. Using the SS-Tool, we corrected 
the raw ramps for non-linearity and converted them to the average current values after 
removing spikes and glitches caused by cosmic-ray events and any other discontinuities. 
The SS-Tool process makes a dark current subtraction using 
the shutter-closed data before and after each observation.

The detector output changes with time depending only on the sky brightness during 
the scan observation. However, the output changes even when the telescope is 
pointed towards almost the same sky. This time varying component was identified as 
being from stray light caused by the Earth's limb emission illuminating the telescope baffle. 
The stray light appears in the case when the Earth avoidance angle of the telescope 
is smaller than 80 degrees. The Earth avoidance angle changes with time depending 
on the orbital phase of the satellite and also with terrestrial season according to 
the inclination of the orbit even when the same field is being observed. The stray light 
appears typically for $\sim150$ s at the begin and end of the observation time. 
At larger Earth avoidance angles stray light appears for $\sim200$ s in the mid of 
the observation time.

Stray light is removed in time domain by high-pass filtering the data with a window 
size of 90 s, which was determined so to remove the slowest signal from the stray 
light but not to lose faster signals caused by astronomical emission. For the slow-scan 
observation, the time domain filter is equivalent to a spatial filter in 
the scan direction. We determined the window size of the filter in the scan direction 
as to be equivalent to the FOV size in the cross-scan direction, to ensure isotropy.

\subsection{Flat fielding}\label{sec:flat fielding}
Flat-fielding, i.e., correction of responsivity inhomogeneity in detector array, 
was carried out by measuring the mean brightness of the scanned area for 
each observation. This method is based on the assumption that the sky brightness 
averaged over the observed area is the same for all array pixels. Accordingly, 
this method is effective especially for dark sky dominated by zodiacal emission, 
whose spatial fluctuation is known to be very small from mid-infrared 
observations; their smoothness in $arcmin$ scales being better than 1$\%$ 
\citep{abraham98}. 

According to the results of previous far-infrared missions, e.g., IRAS and COBE, 
the sky brightness towards high galactic latitudes in the shorter AKARI bands 
at 65 and 90 $\micron$ should be dominated by zodiacal emission even at high 
ecliptic latitude; the expected fractions are 90\% at 65$\micron$ and 70\% at 90$\micron$, 
while at 140 and 160$\micron$, Galactic cirrus emission is more likely the major 
($>$50\%) component in many cases. Since the smoothness of Galactic cirrus 
depends on the angular scale and mean brightness, the flat fielding in cirrus 
regions is not so excellent as the method using the zodiacal emission. 
If we choose a sky with a brightness of about 10 MJy sr$^{-1}$, the smoothness 
of the cirrus can be better than 10\%.

In order to construct a flat-field template for the short wavelength bands, we have 
carried out observations of zodiacal emission at ecliptic latitudes lower than 
20 deg in the performance verification (PV) phase. For the long wavelength bands, 
we have also taken cirrus observations at intermediate galactic latitudes of about 
20 deg. However, the use of such a pre-measured flat has suffered from 
the responsivity change induced by radiation effects and the degradation 
of detector properties, which may not be perfectly corrected. 

For the ADF-S survey, in fact, flat fielding with the observed sky itself has 
yielded much better results than the pre-measured flat, because the responsivity 
change could be effectively corrected by the calibration sequence made 
during each observation. Thus, we constructed such "self-flat" data by averaging 
the time series data for each pixel after a 3-sigma rejection in time during 
the settling time for attitude control and the initial scan observation. The flat-field 
accuracy with this method was generally better than 5\%, and the actual effect 
of the flat field error for each observation was taken into account in the noise 
map derived from the signal variance for multiple scans over the same sky. 

\subsection{Absolute calibration}\label{sec:absolute calibration}
\subsubsection{Pre-flight measurement in laboratory}
Pre-flight measurements of the absolute gain including the detector responsivity 
were carried out in all the wave bands of AKARI in the laboratory, using an external 
blackbody source at various temperatures ranging from 17K to 30K with attenuators 
of 20-30dB. The measured gain factors show good agreement with each other to 
within $\pm$5\%, and we adopted a mean value as the final gain factor. 

In order to compare any measurements done at different times and different 
conditions either in orbit or in the laboratory, we have corrected the absolute 
gain value by referring to internal calibrator signals. The reproducibility 
of the calibrator signals was approximately $\pm$5\%, and this was also 
the limiting factor of the calibration accuracy. In the short wavelength bands at 
60 and 90 $\micron$, we estimated the accuracy of the pre-flight calibration 
to better than $\pm$7\%, by combining the above uncertainties quadratically.

While the monolithic Ge:Ga detectors for the short wavelength bands had 
fairly good uniformity of the responsivity over the array and showed a high 
stability in orbit, the stressed Ge:Ga detector arrays for the long wavelength 
bands showed considerably larger responsivity differences among pixels and 
strong radiation effects in orbit. These properties made it difficult to correct for 
the responsivity in the long wavelength bands even with the internal calibrator 
(c.f. flat field error) and caused additional calibration errors of $\pm$6\% and 
$\pm$9\% at 140 and 160 $\micron$, respectively.

\subsubsection{In-orbit measurement}
In order to measure the absolute gain of the detectors in orbit, we carried out 
observations of diffuse emission whose brightness had been previously accurately 
measured with COBE, as described in the last section. 

For the short wavelength bands, we observed the zodiacal emission at low 
ecliptic latitudes. To derive the gain factor, we compared our observed data with 
the COBE/DIRBE model of zodiacal emission, which is reliable at low ecliptic latitude 
to an accuracy better than 2\% \citep{kelsall98}. In order to use this method avoiding 
the contamination from Galactic and extragalactic emission, we differentiated 
the sky brightness measured at different ecliptic latitudes in low cirrus regions 
with similar HI column densities (N$_{HI}\sim$2$\times$10$^{19}$ cm$^{-2}$) 
and compared the differential data with the zodiacal light model. The accuracy 
of this method is consequently limited by the calibration uncertainty of DIRBE 
 $\sim$10\% \citep{hauser98}.

For the long wavelength bands, we observed Galactic cirrus and compared 
the measured brightness with the COBE/DIRBE annual average map (AAM). 
The sky brightness at low galactic latitudes in this wavelength range is dominated 
by Galactic cirrus, seasonal variation of zodiacal emission is smaller than 3\%, 
and the AAM map is thus valid for this calibration. In addition to the calibration 
uncertainty of DIRBE, the low-frequency noise and instability of the detector 
during the observations increased the uncertainties to $\pm$15\% and  $\pm$19\% 
at 140 and 160 $\micron$, respectively.

\subsubsection{Final gain factor and uncertainty}
For the short wavelength bands, the absolute gain factor of AKARI from pre-flight 
measurements showed excellent agreement with that from the in-orbit calibration. 
Their difference was only 5\% at both 65 and 90 $\micron$. It is noteworthy that 
this difference is smaller than either of the uncertainties of the pre-flight calibration 
and the in-orbit calibration. For the long wavelength bands, the pre-flight measurement 
gave a slightly different gain from that of the in-orbit calibration; the difference was 
13\% and 25\% at 140 and 160 $\micron$, respectively, but they agreed with each 
other within the uncertainties.

In this work, we adopted our own gain factor from the pre-flight measurement 
to indicate the final result of the CIB brightness. For the foreground subtraction, 
however, we adopted the in-orbit calibration factor to compare our data directly 
with the DIRBE result. In summary, the absolute calibration uncertainties of our 
CIB measurements taking all systematics into account are $\pm$7\% at both 
65 and 90 $\micron$$, \pm$9\% at 140 $\micron$, and $\pm$13\% at 160 $\micron$. 
These numbers are sufficiently small enough for any significant detection of the CIB. 
More details about the absolute calibration are described in a separate forthcoming 
paper (S. Matsuura et al. 2010, in preparation).

\subsection{Co-added image}\label{sec:co-added image}
Each slow-scan observation covers only a small patch of the sky corresponding 
to the scanned area. To produce the final mosaic image of the entire ADF-S 
field from the multiple scans, we co-added the time-series data in grid cells of 
the sky coordinates, according to the sky position of each detector pixel at 
each time of the data sampling. The co-addition process was done after 
subtracting the zodiacal foreground, which showed a seasonal variation as 
described in the next section. 

Even after subtraction of the zodiacal foreground, the mosaic image shows 
stripes as a scanning effect due to imperfect correction of the detector responsivity. 
In order to de-stripe the image, we determined the baseline levels of 
the zodi-subtracted images taken in the same season by a smooth filtering 
in the ecliptic longitude (cross-scan) direction with a filter width of 30 arcmin, 
which corresponds to three times the scan width. This filter works as 
a notch filter to reduce the noise only at the angular scale of the scan 
width preserving the fluctuation power at other angular scales. The 
 baseline levels are then subtracted from the zodi-subtracted time-series data. 

Finally, we produced the co-added image by calculating the weighted mean 
after a 3$\sigma$ rejection in each grid cell of the image. We estimated 
the standard error of the mean value from the standard deviation and number 
of data samples. The size of the grid cell is determined so as to contain 
sufficient number of data samples, according to the data interval, the scan speed, 
number of the scans, and number of the array pixels seeing the same sky. 
For the ADF-S survey, the grid cell size was set to be 20 arcsec $\times$ 20 
arcsec each containing more than 8 data samples on average.

\section{Foreground subtraction}\label{sec:foreground subtraction}
In order to produce the final mosaic image of the entire field by the co-addition 
of the time series data, we have adjusted the mean sky levels of multiple observations 
by subtracting the zodiacal foreground with its seasonal variation. For accurate 
measurements of the CIB, estimates of the contribution from Galactic cirrus and 
external galaxies are also essential. In this section, we describe our methods 
of the foreground subtraction and show the final mosaic image of the entire 
ADF-S field. 

\subsection{Zodiacal emission}\label{sec:zodiacal emission}
\subsubsection{Seasonal variation of zodiacal brightness}
Although the seasonal variation of the zodiacal brightness is sometimes 
an obstacle, it is rather useful to know the contribution of dust near 
the Earth to the zodiacal brightness. In fact, conventional models of the zodiacal 
cloud on the basis of the IRAS and COBE observations have been constructed 
relying on the local dust density and temperature derived from the seasonal 
variation \citep{reach88,kelsall98}. The amplitude of the seasonal variation is not 
affected by any uncertainty from the isotropic component of the zodiacal emission.

In Figure~\ref{fig:time_var65}, the measured sky brightness at 65 $\micron$ 
at various locations in the ADF-S are plotted as a function of days after launch. 
Figure~\ref{fig:time_var90} is the same plot but at 90 $\micron$. The sky brightness 
values are only for those dates when the field were observed, rather than 
the whole length of the mission. Statistical errors of the sky brightness from 
the RMS fluctuation of the image are of the order of a few kJy sr$^{-1}$ 
(smaller than the size of the plot symbols). The sky brightness in the second 
season in Winter, about a year after launch, is significantly lower than that in 
the first and third seasons in Summer, as expected from a one-year period of 
the seasonal variation of zodiacal brightness. 

For comparison, the sky brightness towards the LH and NEP, which have been 
measured as supplemental observations in previously well explored fields, 
are also shown in Figure~\ref{fig:time_var65} and Figure~\ref{fig:time_var90}. 
The LH for both PV and MP are brighter than the ADF-S because of the fields 
higher zodiacal brightness in spite of the similar density of Galactic cirrus. 
The NEP is brighter than the ADF-S due to the larger contribution from Galactic 
cirrus even though it has the lowest zodiacal brightness. 

\begin{figure}[!ht]
\begin{center}
   \resizebox{1.0\hsize}{!}{
     \includegraphics*[0,0][900,600]{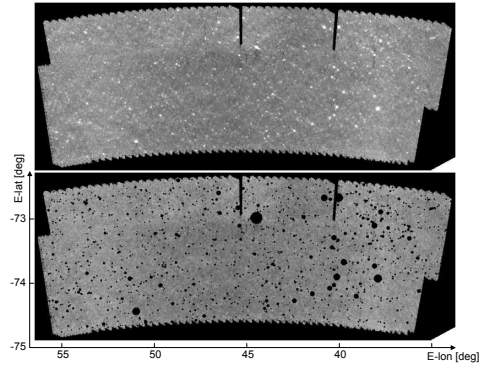}
   }
\caption{
Upper: Diffuse map of ADF-S at 90 $\micron$ after subtraction of the zodiacal 
foreground, drawn as linear contours. Brighter the contour, higher the surface 
brightness. The map size is approximately 6 deg $\times$ 2 deg. 
Lower: The source removed map after masking out the sources down to 
$\sim$20 mJy. The masked area is filled with zeros to show the source 
locations clearly.
}\label{fig:image}
\end{center}
\end{figure}

\subsubsection{Residual brightness after zodi-subtraction}
To measure the absolute brightness of the CIB, we subtracted the zodiacal 
foreground for each observation by using the same COBE/DIRBE model as 
we used for the in-orbit calibration. The zodiacal brightness in the AKARI 
bands were calculated by interpolation from the DIRBE bands with the color 
correction assuming a blackbody spectrum. The zodi-subtracted data are 
composed of just the Galactic cirrus and CIB, as long as the zodi-model is correct. 

In Figure~\ref{fig:time_var65} and Figure~\ref{fig:time_var90}, residuals 
after subtraction of the zodiacal foreground are compared with the raw data. 
The zodi-subtracted data for the ADF-S taken in different seasons show good 
agreement with each other, not only in the the mean levels but also 
in the detailed structure mainly due to Galactic cirrus. In addition, the residuals 
for the ADF-S and LH agree with each other except for the detailed structure, 
while the NEP is much brighter than the others, especially at 90 $\micron$ due 
to the larger contribution from the Galactic cirrus, as described later. At 65 $\micron$ 
the difference of the NEP brightness to the other fields is not so clear 
since the sky brightness is dominated by the zodiacal emission. 
The results indicate that the zodi-model is correct to an accuracy better 
than 5\%, which is a typical value of the model uncertainty \citep{kelsall98}. 

\subsubsection{Zodi-subtracted diffuse map}
The upper image of Figure~\ref{fig:image} shows the final diffuse map at 
90 $\micron$ produced by mosaicing the zodi-subtracted data. To obtain this 
mosaic image,  destriping was made by first calculating the mean sky levels 
of the image strips taken by multiple scans after the removal of bright sources 
by a 3-sigma rejection method, followed by time-domain smooth filtering to 
the mean sky levels for the multiple scans to correct any slight responsivity 
difference that may not have been perfectly corrected by the calibration sequence. 
Finally all images are co-added. The window size used for the time-domain filter 
corresponds to three scan stripes in the space domain.  A large contiguous area 
was successfully mapped, except for small lost data areas such as the inlets 
in the upper part of the map caused by problems with the ground station. 
As clearly seen in the map, there exists large scale structure from Galactic cirrus 
emission; with the central region darker than the surrounding area. Most of 
the bright spots can be attributed to nearby galaxies. 

\subsection{Stars and galaxies}\label{sec:stars and galaxies}
Each star or galaxy in the field does not contribute much to the overall sky brightness, 
but the total contribution including faint sources down to the detection limit is 
non-negligible for accurate measurements of the CIB. Moreover, point sources 
seriously affect the sky fluctuations rather more than the absolute brightness. 
Hence, we masked out foreground stars and galaxies down to the $3\sigma$ 
detection limit of $\sim$100 mJy at 65 $\micron$ and $\sim$20 mJy at 90 $\micron$, 
from the diffuse map. The mask is a circular area around each source with 
the radius normalized to be 1 arcmin at 100 mJy increasing proportionally 
with increasing flux. The lower image of Figure~\ref{fig:image} is the resultant 90 $\micron$ 
map after the source removal. The masked areas are filled with zero values to 
make the source locations clear.

According to the point spread function (PSF) previously measured for stars 
and asteroids as calibration standards \citep{shirahata09}, the remaining 
brightness in the out-of-mask region around a point source is less than 5\% 
of the peak brightness, and the integrated flux in the out-of-mask region is 20\% 
of the total flux. The integrated flux of the out-of-mask region of all sources 
down to the $3\sigma$ detection limit is less than half of the total source flux. 
Consequently, the mean brightness of zodi-subtracted map after removing 
the sources in such way is $\sim3\%$ lower than the original value including 
the sources, and the sky fluctuation power is also an order of magnitude lower. 
We have also confirmed that the fluctuation power has the same value even 
if we change the mask aperture by a factor of $\pm50\%$. This implies 
the contribution of the out-of-mask fringes of PSF to the total sky fluctuations 
is negligible. 

\begin{figure}[!ht]
\begin{center}
   \resizebox{1.0\hsize}{!}{
     \includegraphics*[0,0][900,600]{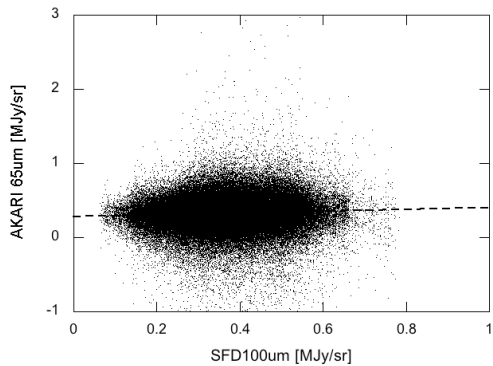}
   }
\caption{
Correlation between the AKARI 65 $\micron$ map and the SFD 100 $\micron$ map. 
The dashed line is the best fit to the data. 
}\label{fig:cirrus65}
\end{center}
\end{figure}

\begin{figure}[!ht]
\begin{center}
   \resizebox{1.0\hsize}{!}{
     \includegraphics*[0,0][900,600]{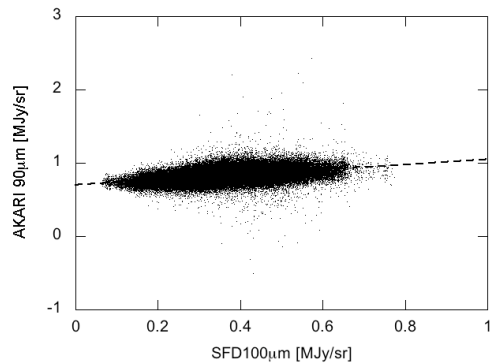}
   }
\caption{
Correlation between the AKARI 90 $\micron$ map and the SFD 100 $\micron$ map, 
same as Figure~\ref{fig:cirrus65}. 
}\label{fig:cirrus}
\end{center}
\end{figure}

\subsection{Galactic cirrus}\label{sec:galactic cirrus}
To estimate the Galactic foreground, we studied the correlation between 
the ADF-S map and the SFD dust map \citep{schlegel98}, which is 
widely used as a reliable map of Galactic cirrus. This correlation may be 
scattered by point sources since the angular resolution of AKARI is much higher 
than that of the SFD map based on the IRAS and COBE data. In fact, many 
bright galaxies in the field are resolved by AKARI, but in the SFD map they 
are left unresolved and recognized as part of the cirrus structure instead. Thus, we 
calculated the correlation for the zodi-subtracted map after removing bright 
galaxies and reducing the map resolution to be the same as SFD.

Figure~\ref{fig:cirrus65} and Figure~\ref{fig:cirrus} show the correlation 
between the SFD map at 100 $\micron$ and the ADF-S map at 65 and 
90 $\micron$, respectively. The 90 $\micron$ map correlates significantly with 
the SFD map with a correlation coefficient of $r$=0.43, and the linear 
coefficient from the best fit line, shown in Figure~\ref{fig:cirrus} by the dashed 
line, is $I(90\mu$m$)/I(100\mu$m$)=0.7030\pm$0.0007. This color ratio 
corresponds to a dust temperature of $T$=18.2 K for $n$=2, where $n$ 
is the emissivity index and $I(\nu)\sim B(T)*\nu^{-n}$, and is consistent 
with typical dust properties \citep{schlegel98}. 
The correlation for the 65 $\micron$ map was marginally seen as $r$=0.05 
with $I(65\mu$m$)/I(100\mu$m$)=0.283\pm$0.002. 

It is remarkable that the zero intercepts of the best-fit lines for the correlations 
for both the 65 and 90 $\micron$ data show significantly finite values, which 
implies a positive detection of an isotropic emission component of the CIB. 
To estimate the cirrus brightness in the long wavelength bands, we extrapolated 
the 90 $\micron$ brightness with a blackbody spectrum of $T$=18.2 K 
for $n$=2, because the map quality was not high enough to allow the same 
correlation analysis as made for the short wavelength bands.

\section{Power spectrum analysis}\label{sec:power spectrum analysis}
\subsection{Method}
In order to carry out a fluctuation study of CIB, we have made a power spectrum 
analysis using the two-dimensional (2-D) FFT for the final zodi-subtracted 
maps of the ADF-S. For the 2-D FFT analysis, we first produced 
a co-added image of the entire field with Cartesian coordinates by a simple 
tangential projection. The image size is so small that the distortion by 
the projection does not affect the result (while a spherical harmonics expansion 
with multipoles is required for power spectrum analysis at angular scales 
larger than 10 $deg$). Next, we subtracted the mean offset level from the image, 
we set the blank data area out of the boundary of the observed area (dark area 
in Figure~\ref{fig:image}) to zero value, and we also set the source-masked 
regions to zero value. Then, we applied a 2-D FFT to the entire image. 
Finally, we corrected the raw power spectrum in the spectral domain by correction 
factors such as effective area, noise spectrum and PSF, as described below.

\subsection{Effective area}
As seen in Figure~\ref{fig:image}, the survey field has a non-rectangular 
shape, and the image includes zero-value data. Such an irregular shape may 
distort the power spectrum, and the non-negligible number of zero-value data points 
decreases the amplitude of the power spectrum. In order to obtain the real power 
spectrum avoiding such artifacts, we have corrected the power spectrum 
by multiplying it by a correction factor corresponding to the effective area of the field.  

To derive this correction factor, we used a random noise simulation method 
by generating noise images of the same shape and area as the survey field, 
so that the root-mean-square noise of all the images have a constant value. 
Then, we calculated the power spectra of the noise images by using the 2-D FFT. 
Since the simulated noise image should have a flat power spectrum, i.e., white noise, 
we can find any effects due to image irregularity as a change in distortion and amplitude 
of the noise power spectra. The correction factor was actually calculated as the ratio 
between the original white noise spectrum and the mean power spectrum of 
the simulated noise images. 

\begin{figure}[!ht]
\begin{center}
   \resizebox{1.0\hsize}{!}{
     \includegraphics*[0,0][900,600]{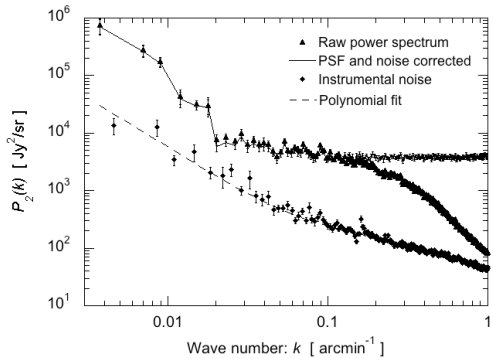}
   }
\caption{
Raw power spectrum of the zodi-subtracted ADF-S map at 90 $\micron$ including 
bright sources (filled triangles), and the PSF corrected power spectrum after 
subtraction of the instrumental noise (thin line). The diamonds represent 
the measured power spectrum of the instrumental noise, and the dotted line 
is the result of a polynomial fit to the noise spectrum. 
}\label{fig:rawpsd}
\end{center}
\end{figure}

\subsection{Angular power spectrum}
Figure~\ref{fig:rawpsd} shows the power spectrum of the 90 $\micron$ map including 
bright galaxies as a function of angular wavenumber. The amplitude is corrected 
by the effective area as described above. To obtain the one-dimensional power 
spectrum, we have projected the 2-D power spectrum onto a radial axis of polar 
coordinates in angular wavenumber by calculating the ensemble mean for phase 
angles in a coaxial ring of each bin of the radial axis. The bin width is approximately 
0.003 arcmin$^{-1}$ corresponding to the inverse of the angular size of the map. 
The error bars of the power spectrum denote the standard error calculated 
from the standard deviation and number of data in each radial bin.

\subsection{Noise estimate}
In order to obtain the fluctuation power with astronomical origin, the subtraction 
of the instrumental noise is necessary. We estimated the noise power spectrum 
from the difference of two subset maps taken in the same field but in different 
seasons. The obtained noise power spectrum shown in Figure~\ref{fig:rawpsd} 
was dominated by the low-frequency noise as fitted by polynomials with terms 
of $1/f$ and $1/f^2$ noise. For the noise subtraction we used the result of 
polynomial fit rather than the data points to avoid adding any artifacts. 

The estimated noise is mainly due to instrumental noise but may include 
systematic errors of the zodi-model uncertainty and the responsivity correction. 
The seasonal variation of small-scale structures of the zodiacal emission may 
also contribute to the noise, but its contribution is also subtracted from 
the measured power spectrum as a noise component. This noise level 
gives an upper limit for the residual fluctuations of zodiacal emission 
smaller than 0.5\% of the mean level at angular scales of $\sim$1 arcmin, 
while the previously reported upper limit at the same angular scale 
is 1\% \citep{abraham98}. 

\begin{figure}[!ht]
\begin{center}
   \resizebox{1.0\hsize}{!}{
     \includegraphics*[0,0][900,600]{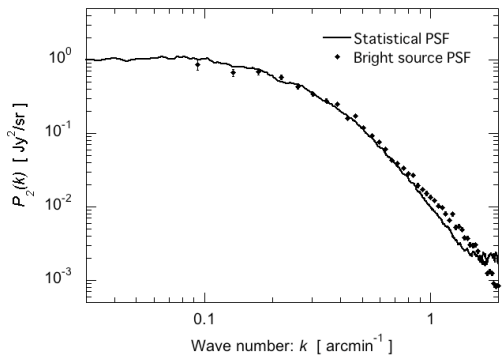}
   }
\caption{
The normalized power spectrum of PSF at 90 $\micron$ directly obtained from 
the ADF-S map itself by two methods. The thin line is the statistical PSF 
obtained by differentiating the maps before and after removing point sources. 
The data points represent the average PSF for bright sources in the field.
}\label{fig:psf}
\end{center}
\end{figure}

\subsection{Beam correction}
The steep decrease of the fluctuation power towards large wave numbers 
is obviously due to low-pass filtering by the PSF, which has a beam size 
of approximately 30 arcsec in FWHM. In order to reproduce the intrinsic 
power spectrum of the map, we divided the raw power spectrum by the normalized 
power spectrum of the PSF constructed from sources in the ADF-S field itself 
(self-measured PSF). Although the pre-measured PSF of bright stars as calibration 
standards could also be used for this purpose, observations of calibration stars 
with the same AOT parameter as the ADF-S survey were not available 
because the calibration sources are too bright. 

Figure~\ref{fig:psf} shows two kinds of power spectra for the self-measured 
PSF; one is a statistical PSF spectrum from the difference between two 
spectra of the source-removed map and the original map, and the other is the 
average PSF spectrum of bright sources only in the field. To obtain 
the average PSF spectrum, we cut out postage stamps with a size of 
40$\times$40 arcmin$^{2}$ around the bright sources with signal-to-noise ratios 
higher than 30. The statistical PSF spectrum is normalized at the small end of 
wavenumber, and the average PSF spectrum is scaled to fit  the overall 
range. The 2 methods show good agreement with each other in the range of 
0.1-0.8 arcmin, though a large discrepancy appears at large wave numbers. 
Since this paper focuses on the power spectrum at large angular scales we have 
adopted the statistical PSF, because it actually provides a better result in order 
to reproduce the power spectrum of the map at large wavenumbers and it is also 
known that transient effects are stronger for brighter sources \citep{shirahata09}.

\begin{figure}[!ht]
\begin{center}
   \resizebox{1.0\hsize}{!}{
     \includegraphics*[0,0][900,600]{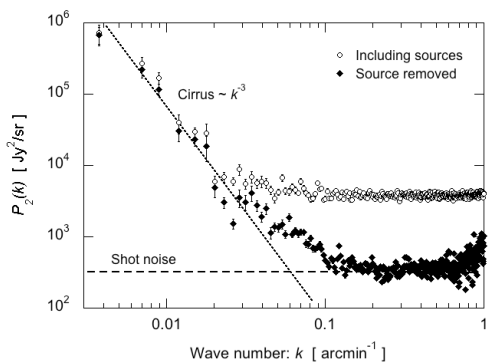}
   }
\caption{
Power spectrum of the ADF-S map after removing point sources (filled diamonds) 
compared with the power spectrum including the sources (open circles). 
The dotted line is the third power law for Galactic cirrus. The dashed line 
represents the shot noise level due to individual contribution of unresolved galaxies.
}\label{fig:realpsd}
\end{center}
\end{figure}

\subsection{Resultant power spectrum}
\subsubsection{Raw power spectrum}
The power spectrum after the noise subtraction and the PSF correction is shown 
in Figure~\ref{fig:rawpsd}. There exist two major components in the power spectrum. 
One is the constant component dominating at wavenumbers larger than 
0.1 arcmin$^{-1}$, and another shows a steep rise towards the smallest wave number 
following a $k^{3}$-power law. The former is regarded as the shot noise (Poisson noise) 
mainly due to bright galaxies in the field, and the latter can be identified as Galactic 
cirrus fluctuations, as discussed in the following section.

We have carried out a power spectrum analysis for the 65 $\micron$ data in 
the same way as the 90 $\micron$ data. However, the obtained power spectrum 
was not significant compared to the noise power, suffering from low sensitivity. 
It was also difficult to obtain a significant power spectrum at 140 and 160 $\micron$ 
because of the presence of low-frequency noise originating from the responsivity 
change induced by cosmic-ray hits. 

\subsubsection{Bright source removal}
The raw power spectrum in Figure~\ref{fig:rawpsd} is obviously contaminated 
by shot noise with a flat spectrum at larger wavenumbers, originating
from randomly distributed bright galaxies in the field. To investigate the fluctuations 
originating from underlying faint galaxies (the building blocks of the CIB), we derived 
the power spectrum removing the point sources down to the detection limit of 
AKARI, $\sim$20 mJy.

Figure~\ref{fig:realpsd} shows the PSF-corrected power spectra before and 
after the point source removal. The shot noise power, in contrast to the cirrus 
component, decreases to $10\%$ of the value before the source removal. 
The increased power at $\sim$1 arcmin$^{-1}$ may be the signature of small-scale 
correlation between galaxies and source confusion or may be due to the uncertainty 
of the PSF as seen in Figure~\ref{fig:psf}. Study of the power spectrum at these small 
angular scales is reserved for future work. At the other wavelengths, we have only 
obtained upper limits for the shot noise, which are not so meaningful as new limits.

\section{Results and Discussion}
\subsection{Mean CIB levels}\label{sec:mean CIB levels}
\subsubsection{Absolute brightness towards the ADF-S}
\begin{figure}[!ht]
\begin{center}
   \resizebox{1.0\hsize}{!}{
     \includegraphics*[0,0][900,600]{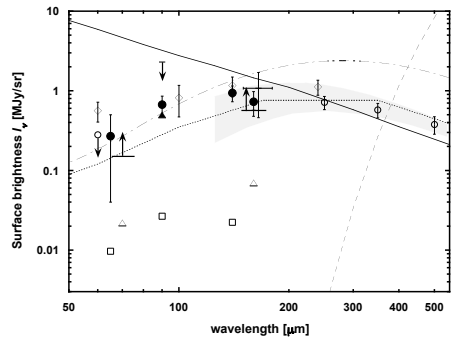}
   }
\caption{
The CIB brightness measured in the ADF-S (filled circles), compared with the 
previous CIB limits from COBE/DIRBE (diamonds), COBE/FIRAS (shaded region), 
IRAS (open circle with downward arrow at 60 $\micron$), ISO (thin line with 
downward arrow at 90 $\micron$ and cross in 150-180  $\micron$), Spitzer 
(upward arrows) and BLAST (open circles at 250, 350 and 500 $\micron$). 
The filled triangle at 90 $\micron$ is the lower limit of the CIB derived for 
the case of "very-strong no-zodi" principle \citep{wright01,levenson07}. 
The solid line shows a typical spectrum of the zodiacal foreground at the ecliptic 
poles. The dashed line is the CMB. Model predictions from \citet{lagache04} and 
\citet{takeuchi10} are indicated by the dotted and dot-dashed curves, respectively. 
The open triangles and the open squares are lower limits obtained from the integrated 
flux of the galaxy counts with Spitzer and AKARI, respectively.
}\label{fig:cirb}
\end{center}
\end{figure}

The resultant CIB brightness was measured as the zero-intercept of the correlation 
between the zodi-subtracted map and the SFD map as shown in Figure~\ref{fig:cirb} 
and summarized in Table~\ref{tbl-2} with statistical and calibration errors with an 
uncertainty in the zodiacal emission model of 5\% \citep{kelsall98}. In Table~\ref{tbl-2} 
the mean brightness of zodiacal emission and Galactic cirrus emission over the entire 
ADF-S field are also shown as components of the measured total sky brightness. 
The measured brightnesses in Table~\ref{tbl-2} are all color-corrected assuming 
a flat spectrum. Although the zero-point offset of the SFD map may cause a 
systematic uncertainty in the CIB brightness, such uncertainty has been estimated 
to be smaller than 0.1 MJy sr$^{-1}$ at 100 $\micron$ \citep{finkbeiner00}, which 
does not significantly alter the results. In conclusion we can say we have significantly 
detected the CIB in the short wavelength bands. We have also tentatively detected 
the CIB at 140 and 160 $\micron$ by extrapolating the cirrus brightness from 
90 $\micron$ to each wavelength assuming a $T$=18.2 K blackbody spectrum.

In order to obtain a more robust result, if we adopt another zodi-model with a very 
powerful assumption about the celestial signal called "very-strong no-zodi" 
principle, i.e., no extrasolar emission at 25 $\micron$ exists at high ecliptic 
latitudes \citep{wright98,wright01,levenson07}, then the zodi-model brightness 
becomes $\sim$10\% higher than what was assumed for our foreground 
subtraction (L. Levenson 2009, private communication). On this assumption, 
the residual background decreases completely to zero at 65 $\micron$ but still 
remains at 90 $\micron$ at a significant positive value of 0.50 MJy sr$^{-1}$, 
because of the bluer color of zodiacal emission. This lower limit of the CIB at 
90 $\micron$ is shown in Figure~\ref{fig:cirb}. In the long wavelength bands, 
the zodi-model difference changes the residual background by at most 10\%, 
which is smaller than the statistical error. 

\subsubsection{Isotropy}
In Table~\ref{tbl-3}, we summarize the zodi-subtracted sky brightness in 
various fields as a mean value with the error estimated from the brightness 
variance for a number of observations in each field. The error does not include 
either the calibration error or the zodi-model uncertainty. The table also 
indicates the residual brightness after subtracting the cirrus brightness, which 
were estimated from the HI 20 cm line map by the Leiden-Argentine-Bonn (LAB) 
survey \citep{kalberla05}. The cirrus brightness was obtained by assuming 
a conversion factor from the HI-column density to the 100 $\micron$ brightness of, 
$0.6\pm0.1$ MJy sr$^{-1}/10^{20}$cm$^{-2}$, as the median value of various 
COBE-based results with their variation as an uncertainty \citep{boulanger96,
reach98,schlegel98,lagache00a}. For the wavelength conversion from 
100 $\micron$ to the AKARI bands, we used the correlation coefficients in 
Figure~\ref{fig:cirrus65} and Figure~\ref{fig:cirrus}. Since the angular resolution 
of the HI data is 10 arcmin, much larger than that of our data, we used the mean 
brightness of our data and of the HI data in each field. The residual brightness 
after subtraction of the cirrus brightness in different fields agree with each other 
within the error bars at both 65 and 90 $\micron$. 

The ratio of the HI-column density to the far-infrared brightness assumed 
above may be an overestimate, because it tends to be determined by the data 
in the brighter cirrus regions. In low cirrus regions the ratio could be smaller due 
to the contribution of uncorrelated components, such as dust associated with 
ionized gas. In fact, the mean residuals for the three fields in Table~\ref{tbl-3} 
are slightly smaller than the results in Table~\ref{tbl-2} from the correlation of 
the ADF-S and SFD maps, though they are consistent with each other within 
the total error including the calibration error. The contribution of cirrus could be 
more accurately estimated in the long wavelength bands, where the cirrus emission 
is brighter. However, we have not made the same analysis as the short wavelength 
bands, because accurate measurement of CIB was possible only in the ADF-S 
field which has highly statistical samples.

\subsubsection{Comparison with previous observations}
As seen in Figure~\ref{fig:cirb}, our results in the four photometric bands show 
good agreement with previous CIB measurements from COBE at 60, 100 and 
140 $\micron$ \citep{hauser98}. They are also consistent with an upper limit from 
IRAS at 60 $\micron$ \citep{miville02} and with an upper limit at 90 $\micron$ and 
the measured CIB level in the 150-180 $\micron$ range measured by ISO 
\citep{juvela09}. 

The integrated galaxy counts \citep{dole04,matsuura07,shirahata08} account for 
only a few per cent of the CIB brightness. Stacking analysis of the Spitzer 24 $\micron$ 
sources \citep{dole06} substantially extended the lower limits at 70 and 160 $\micron$ 
to the prediction from the galaxy evolution model presented in \citet{lagache04}. 
Our measured CIB brightness at 90 $\micron$ is obviously higher than the Spitzer 
result, and our result implies the existence of large number of unknown sources, 
which may have been missed in the stacking process since they were too faint to 
be detected at 24 $\micron$ with Spitzer. 

A hypothetical source with such an excess brightness should have peculiar 
spectra with a peak around 100 $\micron$, so that they have only a small contribution 
to the CIB in the other wave bands. High levels of the CIB at 100 $\micron$ have 
already been seen with COBE \citep{hauser98,finkbeiner00} but have been 
thought to be due to calibration errors and foreground uncertainties \citep{dole06}. 
It is remarkable that this work corroborates with the COBE result using a different 
instrument.

\subsubsection{Possible origins of the excess emission}
If the excess emission at 90 $\micron$ originates from an isotropic component 
of the Galactic cirrus, which may not be accounted for in the SFD map, it should 
be tightly constrained by the previously measured HI-column density unless 
the gas-to-dust ratio is extraordinary. If the excess brightness at 90 $\micron$ of 
$\sim$0.4 MJy sr$^{-1}$ is associated with neutral Hydrogen, the corresponding 
HI-column density must be $N_{HI}$$\sim1\times10^{20}$ cm$^{-2}$. This is 
much higher than the measured HI-column density towards the ADF-S 
(see Table~\ref{tbl-3}) and unlikely, judging from uncertainty of the HI observation. 
Moreover, if the temperature of such isotropic dust emission is $\sim$17 K, similar 
to the bulk of the Galactic cirrus emission, it should appear at longer wavelengths 
as a strong excess, but this is not significant in our data. Therefore, it is difficult to 
explain the excess brightness by dust emission associated with neutral Hydrogen.

Unexpectedly large amounts of ionized gas at high galactic latitudes as a warm 
ionized medium (WIM) is a another possible candidate for the excess emission. 
In fact, dust emission from a WIM has been detected with the WHAM 
H$_{\alpha}$ survey, though the observed areas have been limited to relatively 
dense regions at low galactic latitudes \citep{lagache00a}. Large differences 
in the residual brightness between different fields shown in Table~\ref{tbl-3} 
can be caused by the contribution of dust associated with a WIM. In fact, the ADF-S 
is darker than the LH in the far-infrared, while the HI-column density towards 
ADF-S is higher than that towards LH. However, any WIM could be dust poor 
compared to the Galactic plane, because grain destruction is expected to 
occur in low density regions of the interstellar medium, while the ionized Hydrogen 
column density to account for all of the excess brightness must be higher than 
that of the neutral Hydrogen. As in the case of the neutral Hydrogen, the dust 
temperature must be so high as to have an emission peak around 100 $\micron$ 
while not producing a significant excess at longer wavelengths, but the existence 
of such hot dust at high galactic latitudes has not been reported. Future 
measurements of the H$_{\alpha}$ emission and other interstellar lines in the
ADF-S would be important tests of the WIM hypothesis to explain the excess.

Star-forming galaxies at high redshift are another possible origin of the CIB 
excess in the far-infrared. In Table~\ref{tbl-2}, the measured CIB brightness 
is compared with the range predicted from a selection of galaxy evolution models, 
which take the contribution of a population of luminous infrared galaxies peaking 
at $z\sim1$ and ultra-luminous infrared galaxies at $z>2$ into consideration 
\citep{lagache04, pearson07, franceschini08, takeuchi10}. Most of these 
galaxy evolution models give a lower brightness than our measured CIB at 
65 and 90 $\micron$, while the models agree with our result at 140 and 160 
$\micron$, the stacking analysis of Spitzer data at 70 and 160 $\micron$, 
and the COBE results at wavelengths other than 100 $\micron$, within 
the data uncertainties. A model prediction from \citet{takeuchi10} in 
Figure~\ref{fig:cirb} shows a fairly good agreement with the measured CIB 
at wavelengths shorter than 140 $\micron$, but it gives too high brightness 
at longer wavelengths, especially in the submillimeter range \citep{puget96,
fixsen98}. It is obvious that our result especially at 90 $\micron$ provides a
tight constraint on galaxy evolution models. 

In contrast to the CIB measurement, the source counts at 90 $\micron$ with 
AKARI were much lower than those predicted from contemporary galaxy evolution 
models, while the galaxy counts at 65 and 140 $\micron$ are consistent with 
the models and the Spitzer results at 70 and 160 $\micron$ \citep{matsuura07,
shirahata08}. Such low counts at 90 $\micron$ have also been found with ISO 
\citep{rodighiero03,heraudeau04}, and it has been reported that many galaxy 
evolution models fail to reproduce the 90 $\micron$ counts \citep{chary01,
serjeant04}. At 90 $\micron$ the integrated galaxy counts down to the detection 
limit of AKARI, $\sim$20 mJy, can only account for less than 10\% of the CIB 
brightness \citep{matsuura07}. To reproduce the CIB brightness, a strong 
increase in the galaxy counts below the detection limit is necessary. If we 
assume a single power law with power index of -1.5 (Euclidian) for 
the cumulative galaxy counts, $dN/dS\sim S^{-1.5}$, the power law needs 
to extend to 0.01 mJy. Such high counts require an extraordinary evolution 
scenario because conventional galaxy evolution models predict a steep drop 
of the differential 90 $\micron$ counts towards the faint end below 1 mJy. 

To explain the previous COBE result of high CIB levels at 60 and 100 $\micron$, 
the possible contribution of a new population at high redshift ($z=2-3$) with 
high dust temperatures ($T_d\sim$60 K) heated by AGN, or accretion onto a
black hole, has already been discussed \citep{finkbeiner00,brain02,brain04}. 
If this is true, our finding of the CIB excess at 90 $\micron$ could also be explained 
by such high-$z$ hot sources. Many results from the Spitzer 24 $\micron$ and 
X-ray surveys suggest that LIRG and ULIRG at $z\sim1$ rather than AGN 
at high redshift constitute main bulk of the CIB, and the contribution of AGN 
to the CIB is thought to be at most 10\% \citep{treister06,devlin09}. 
Therefore, the CIB excess is more likely to be composed of high-$z$ hot 
sources of new galaxy population rather than known obscured AGN, which 
could be detected at 24 $\micron$. If the hot source is so heavily obscured that 
the dust extinction is strong even in the mid-infrared, and if the dust grains are so 
small that the emissivity index is a maximum, $\lambda^{-2}$, at longer wavelengths, 
they can contribute to the CIB selectively at short far-infrared wavelengths.

\subsubsection{Energetics and cosmological implications}
The total integrated energy of the CIB excess in a range of the AKARI bands 
is estimated to $\sim$10 nWm$^{2}$sr$^{-1}$, which is comparable to the CIB 
energy associated with the 24$\micron$ galaxies of 24 nWm$^{2}$sr$^{-1}$ 
\citep{dole06}. If we assume a constant star formation rate (SFR) over 
cosmic history of $z>1$, the excess energy corresponds to the SFR of 
$\sim0.1$ M$_{\sun}$yr$^{-1}$Mpc$^{-3}$, which is an order of magnitude 
higher than that in local universe \citep{hauser01}. If the radiation energy 
is produced by a star burst at $z$, the total hydrogen mass fraction converted 
to metals in this epoch is estimated to $\Delta X\sim0.004(1+z)$. For redshift 
of $z>1$, more than 1\% of baryon mass density needs to be converted to 
metals. Starburst at high-$z$ as the origin of the CIB excess may cause 
over production of metals beyond the metal abundance in local universe. 
Black hole accretion may pray important roll to generate the background 
radiation at high redshift \citep{finkbeiner00}.

In the near-infrared, the CIB excess which cannot be explained by the integrated 
flux of known galaxy populations has been reported, and the origin of the excess 
as first stars, or proto-galaxies, at $z\sim10$ has been discussed in terms of 
the spectral signature of the redshifted Ly-$\alpha$ in the CIB near 1 $\micron$, 
as seen in Figure~\ref{fig:cib_intro} \citep{kashlinsky05,matsumoto05}. 
Findings of the CIB excess in both near- and far-infrared lead us to consider 
their connection; the UV light from the first massive stars which produce 
the near-infrared excess might be partly absorbed by the self-produced dust, 
re-emitted in the mid-infrared and redshifted into the far-infrared band at present. 
Accretion to the first blackhole as a remnant of the first star explosion might 
also be the hot source at high redshift. Dust production by the first stars and 
successive process as blackhole formation and accretion is little understood, 
and this research subject would be proceeded by future observations of infrared 
galaxies and the CIB in many details.

\subsection{Fluctuations of the CIB}\label{sec:fluctuations}
\subsubsection{Galactic cirrus}
In order to obtain information on the CIB fluctuations from the measured 
power spectrum, it is necessary to estimate the fluctuation power of 
the foreground such as the zodiacal emission and the Galactic cirrus. Zodiacal 
emission, the most dominant foreground component, has been known to 
have a very smooth distribution, and its fluctuations in the mid-infrared have 
been measured with ISO to be smaller than 1\% of the mean brightness 
at angular scales of 1 arcmin \citep{abraham98}. This value is already much 
smaller than the shot noise level by unresolved galaxies. Although an 
unexpected time varying residual of zodiacal emission after the zodi-model 
subtraction may contribute to the power spectrum over a broad range of angular 
scales, it has already been subtracted as a part of the noise power estimated 
from the difference of the two sub-maps, as described in the last section. 
Thus, the main foreground is Galactic cirrus, especially at large angular scales.

A $k^{3}$-power law of the fluctuation spectrum for Galactic cirrus has 
been reported by many authors, e.g., \citet{miville07}, but it is 
remarkable that in this work the same power law was confirmed at such 
very low flux levels at 90 $\micron$. The power spectrum of the cirrus fluctuations 
can be expressed as 
\begin{equation}
P_{cirrus}(k)=P_0(k/k_0)^{-\gamma},
\end{equation}
where $k$ is the angular wavenumber in arcmin$^{-1}$, and $P_0$ is the power 
normalized at $k_0=0.01$ arcmin$^{-1}$. $\gamma$ is the power law index, 
ranging from 2.5 to 3.1 and is known to be close to 3 for the optically thin case 
at low cirrus regions \citep{schlegel98,jeong05,miville07}. 
As shown in Table~\ref{tbl-4}, our result gives the cirrus power at 90 $\micron$ 
of $P_0=(5.5\pm0.5)\times10^4$ Jy$^2$ sr$^{-1}$, and the power at 
100 $\micron$ is derived to be $P_0=(1.1\pm0.1)\times10^5$ Jy$^2$ sr$^{-1}$ 
from the correlation coefficient between the ADF-S and SFD maps.

At high values of $k$ the power law index may not always be the same as the value 
derived at low $k$, where the cirrus component dominates the total 
fluctuation power. In previous work with the ISO observations, it has been 
proved that a single power law for the cirrus power spectrum is valid up to 
$k\sim$0.3 arcmin$^{-1}$, though this result is certain only in relatively bright 
regions. Since there is no strong reason to change the power law index at 
higher $k$, we assume a single power law over the entire wavenumber range 
for the estimate of the cirrus power.

It has been reported that the fluctuation power of the cirrus emission scales with 
the mean brightness as a power law with index $2-3$ depending on 
the brightness. According to \citet{miville07}, the cirrus fluctuation power at 
100 $\micron$ in the low brightness regime ($B<10$ MJy sr$^{-1}$) is given by 
\begin{equation}
P_0=2.7\times10^6 B^{2.0} Jy^2 sr^{-1},
\end{equation}
where $B$ is the mean brightness at 100 $\micron$. If this is the case, our 
measured power of the cirrus fluctuations corresponds to the mean brightness 
at 90 $\micron$ of $0.14$ MJy sr$^{-1}$, which is slightly lower than the value 
in Table~\ref{tbl-2}. The relation between the fluctuation power and the mean 
brightness may not be valid under such low cirrus conditions. The result 
implies that it is dangerous to estimate the mean CIB brightness from 
the fluctuation power.

\subsubsection{Shot noise of unresolved galaxies}
We have derived the shot noise of unresolved galaxies from the source
-removed power spectrum in Figure~\ref{fig:realpsd} as the mean power 
level in a range of 0.2-0.8 arcmin$^{-1}$, where the PSF is relatively 
accurate, to be $P_{shot}=(3.6\pm0.2)\times10^2$ Jy$^2$ sr$^{-1}$ 
at 90 $\micron$. In Table~\ref{tbl-4} we summarize our result for 
the foreground estimates at small angular scales, where the shot 
noise dominates, into known emission components. The fluctuations 
of zodiacal emission was estimated to be negligible from the discussion 
in the last section. The cirrus component following a $k^{3}$-power law 
is negligible at such small angular scales ($k>0.2$ arcmin$^{-1}$). 
Therefore, the remaining shot noise can be attributed to the fluctuations 
from unresolved galaxies. 

In Table~\ref{tbl-4} we compare our result with the previous work at 90 $\micron$ 
with ISOPHOT \citep{matsuhara00} and with a range of the shot noise levels 
predicted from various galaxy evolution models. Since the detection limit of 
AKARI for point sources is better than that of ISO, the AKARI result shows 
obviously lower fluctuation power than the ISO result. We have estimated 
the shot noise level for each galaxy evolution model by simply integrating 
the squared flux, $S^2$, with the differential source counts, $dN/dS$, as 
\begin{equation}
P_{shot}=\int{S^2(dN/dS)dS}
\end{equation}
for fluxes below 20 mJy. The measured shot noise power is 3-5 times 
lower than any of the model predictions clearly due to the high source counts
of the models near the detection limit. The result implies that the main bulk of 
the CIB at 90 $\micron$ consists of sources much fainter than the detection limit. 

Since the shot noise is the squared flux integrated with the source counts, 
it is more sensitive to the number density of bright sources than the mean 
CIB level, that is  a linear integration of the flux. As described in the last section, 
the mean CIB level provides constraints on the source counts at the very faint end 
below the point-source detection limit. In contrast, the shot noise can provide 
constraints on the source counts near the detection limit. 

The measured shot noise level can be reproduced by the measured source 
counts extended down to 1 mJy with a Euclidian power law (a constant for the
differential counts). In such a simple scenario, sources fainter than 1 mJy 
have little contribution to the fluctuations but become the main contributor 
to the mean brightness. In order to resolve the bulk of the CIB into point sources, 
future infrared telescope with the detection limits much better than 1 mJy at 
$\sim100$ $\micron$ would be required.

\begin{figure}[!ht]
\begin{center}
   \resizebox{1.0\hsize}{!}{
     \includegraphics*[0,0][900,600]{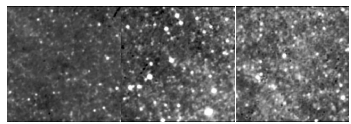}
   }
\caption{
Sub-maps cut out from the selected positions in the ADF-S map at 90 $\micron$; 
left: center region with lowest brightness, mid: a cluster of galaxies (Abell s0463), 
and right: relatively high cirrus region. The image sizes for all maps are the same, 
1$\times$1 deg$^{2}$. 
}\label{fig:submaps}
\end{center}
\end{figure}

\begin{figure}[!ht]
\begin{center}
   \resizebox{1.0\hsize}{!}{
     \includegraphics*[0,0][900,600]{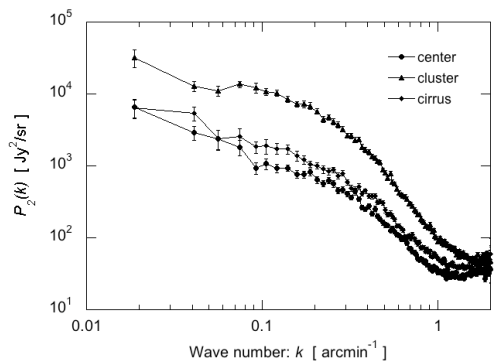}
   }
\caption{
Power spectra of the three maps in Figure~\ref{fig:submaps}. Circles, triangles and 
diamonds represent data points for the central, the cluster of galaxies, and the high 
cirrus regions, respectively. These power spectra are raw data without noise 
subtraction and the PSF correction.
}\label{fig:pdep}
\end{center}
\end{figure}

\begin{figure}[!ht]
\begin{center}
   \resizebox{1.0\hsize}{!}{
     \includegraphics*[0,0][900,600]{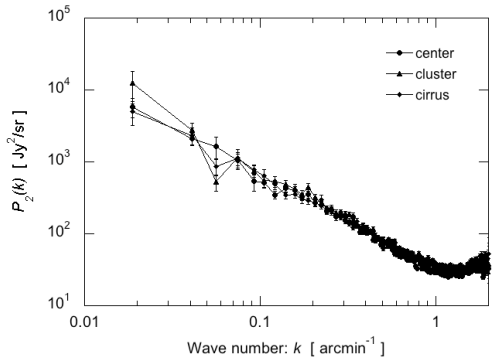}
   }
\caption{
Power spectra for the sub-maps same as in Figure~\ref{fig:pdep} after the point 
source removal.
}\label{fig:pdepsrmv}
\end{center}
\end{figure}

\subsubsection{Cosmic variance}
Since the ADF-S covers a large contiguous area of 12 sq.deg, the field contains 
various regions of different source and cirrus densities. The power spectrum 
shown in Figure~\ref{fig:realpsd} gives the average properties of the entire field, 
but it can be biased due to strong features of specific regions in the map. 

In order to check for any cosmic variance, we have carried out power spectrum 
analysis for sub maps of 1$\times$1 deg$^{2}$ area cut out from various 
positions as shown in Figure~\ref{fig:submaps}. One is the central region 
with the minimum cirrus level and source densities, the second region contains a 
rich cluster of galaxies DC 0428-53 (Abell S0463) \citep{dressler80}, and the
third is a relatively high cirrus region with a moderate source density. 
As shown in Figure~\ref{fig:pdep}, the power spectrum obtained for 
the cluster region is about an order of magnitude higher than the others. 
The central region shows the minimum fluctuation power as expected, and 
the high cirrus region has slightly higher power than the central region, 
though the point source contribution seems to be more prominant rather than 
the cirrus fluctuations. This result implies that the power spectrum is 
severely affected by the point source contribution and biased to high 
source density regions. 

In order to estimate the effect of point sources on the power spectrum in detail, 
we have made the same comparison of the three sub maps after removing 
point sources. The result is shown in Figure~\ref{fig:pdepsrmv}. 
All of the power spectra at the different sky positions show excellent agreement 
with each other at an accuracy better than 5\% in the overall range of angular 
wavenumber. This result implies that the power spectrum after removing 
point sources is not affected by cosmic variance such as the clustering of nearby 
galaxies and can be regarded as a universal function for the CIB fluctuations.

\subsubsection{Clustering component}
In the measured power spectrum in Figure~\ref{fig:realpsd}, a fluctuation 
component of $\sim10^{3}$ Jy$^{2}$ sr$^{-1}$ is seen in the range of 0.03-0.1 
arcmin$^{-1}$. Since the $k$-dependence of this component is moderate
and apparently less steep than the $k^3$-power law, it is not likely to be 
attributed to either shot noise of unresolved galaxies or Galactic cirrus. 
No excess fluctuations of Galactic cirrus on such a small scale have been 
found in the high cirrus density region, where the cirrus dominates the sky 
fluctuations. 

\begin{figure}[!ht]
\begin{center}
   \resizebox{1.0\hsize}{!}{
     \includegraphics*[0,0][900,600]{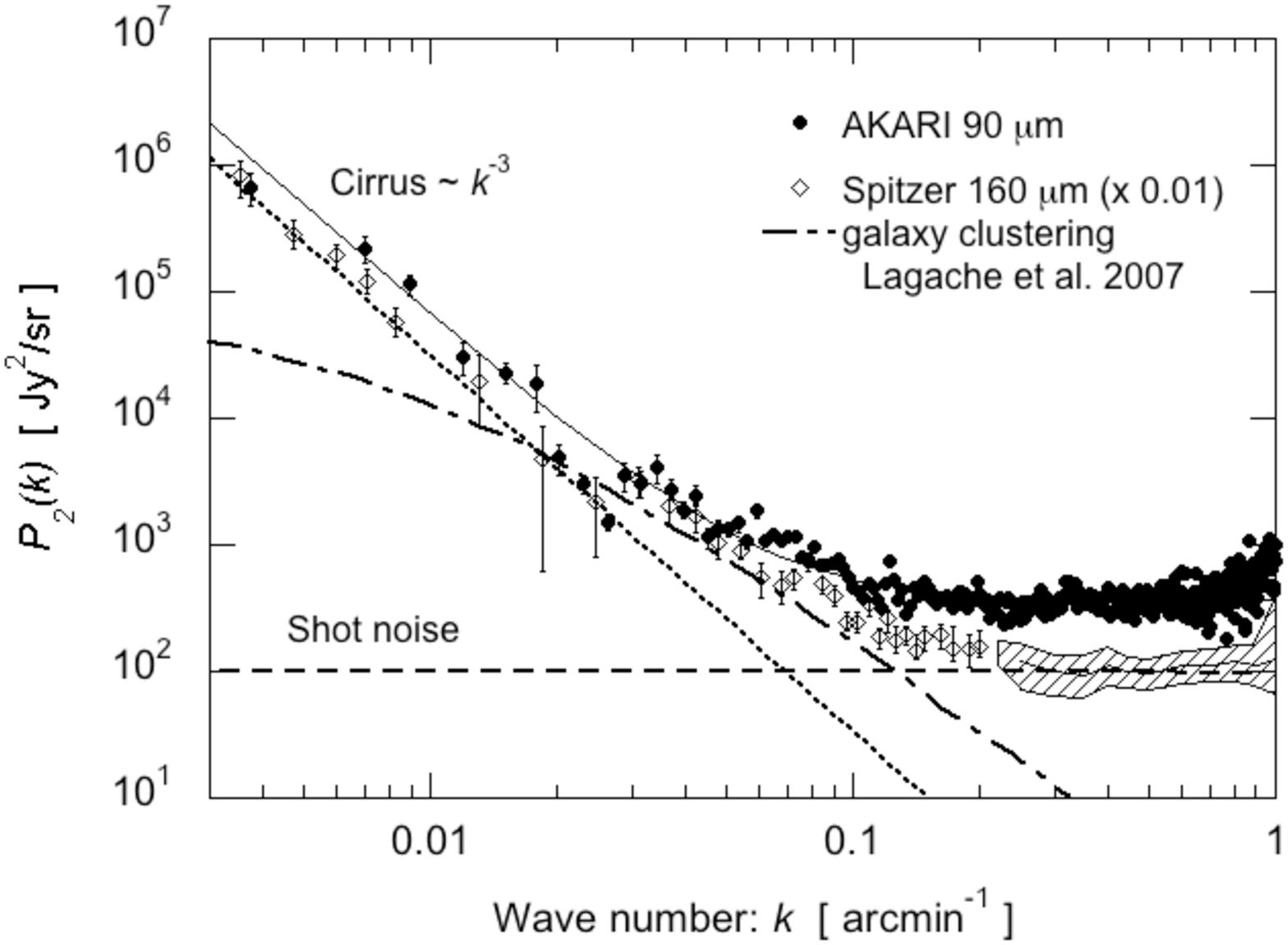}
   }
\caption{
The final power spectrum at 90 $\micron$ (filled circles), same as in 
Figure~\ref{fig:realpsd}, is compared with the Spitzer/MIPS 160 $\micron$ 
result in \citet{lagache07} (open diamonds and a shaded region 
at $k>0.2$ arcmin$^{-1}$) and their model of galaxy clustering 
(dash-dotted curve), Galactic cirrus (dotted curve) and shot noise 
(dashed line). The Spitzer result is multiplied by a factor of 0.01 
to adjust the vertical scale. The solid thin line is the best fit to the AKARI 
data with a multiple power-law model (see text).
}\label{fig:psdmips}
\end{center}
\end{figure}

A plausible interpretation of the fluctuation power at intermediate angular 
scale is the clustering of star-forming galaxies, e.g., \citet{perrotta03}. 
\citet{lagache07} have reported their finding of galaxy clustering in the power 
spectrum at 160 $\micron$ with Spitzer after removing point sources 
brighter than 200 mJy. In Figure~\ref{fig:psdmips}, we plot the Spitzer 
result from \citet{lagache07} multiplied by a scaling factor of 0.01 with our 
power spectrum. Their power spectrum shows an excess component in the
0.01-0.1 arcmin$^{-1}$ range that cannot be explained by the shot noise of 
individual galaxies and the power-law spectrum of Galactic cirrus, agreeing with 
our result. They compared a galaxy clustering model (the dot-dashded 
line in Figure~\ref{fig:psdmips}) with the measured spectrum and directly 
derived the bias factor for structure formation in a dark matter universe 
to be $b\sim2.4$. 

As a result of the comparison of our power spectrum with the Spitzer result, 
the two spectra in $k<0.1$ arcmin$^{-1}$ agree with each other surprisingly 
well not only in the overall shape of the power law component at very low 
frequencies by the cirrus and of the galaxy clustering component but also 
in details such as wiggles at $k=0.01-0.1$ arcmin$^{-1}$ and a dip around 
$k\sim0.02$ arcmin$^{-1}$, which may show an oscillatory structure of the baryonic 
matter distribution. Such concordance between the results of two independent 
observations with different instruments and in different fields leads us to 
the conclusion that the clustering components in both AKARI and Spitzer 
power spectra have the same origin. 

It has been reported that galaxy clustering similar to the AKARI and 
Spitzer results was clearly found by the BLAST experiment in the submillimeter 
range \citep{viero09}. Since the BLAST bands are on the Rayleigh-Jeans side of 
the dust emission in the rest-frame galaxy, BLAST can be more sensitive to high 
redshift galaxies than those of AKARI and Spitzer with less contamination 
by Galactic cirrus. The power spectrum analysis for the BLAST map of 6 $deg^{2}$ 
area,  found a strong signature of galaxy clustering with the power spectrum showing 
a $k^{2}$-power law over the range of $k\sim0.03-0.1$ arcmin$^{-1}$ in all 
the BLAST bands. Their power spectra show a very similar shape to those measured 
by AKARI and Spitzer, in terms of the power-law index of the clustering component 
and the angular scale at the knee between the shot noise and the clustering 
components, even though the limiting flux for the source removal for these three 
missions are different; AKARI $\sim$20 mJy, Spitzer $\sim$200 mJy, and BLAST 
$\sim$500 mJy. 

To extract the clustering component and to compare the AKARI result with 
the results of other missions on the same basis, we have also made a model 
fit to our power spectrum using a $k^{-2}$-power law for the clustering 
component in addition to the shot noise and the cirrus fluctuations, i.e., 
\begin{equation}
P_{2}(k)=P_{cirrus}+P_{shot}+P_{cluster} , 
\end{equation}
and 
\begin{equation}
P_{cluster}=P_{c}(k/k_{c})^{-2} , 
\end{equation}
where $k_{c}$=0.03 arcmin$^{-1}$. Although such a simple model cannot 
reproduce the complex structure seen in the measured spectrum as described 
above. For the best-fit case shown by the thin solid line in Figure~\ref{fig:psdmips}, 
the amplitude of clustering component is 
$P_{c}$=(1.4$\pm$0.8)$\times$$10^{3}$ Jy$^{2}$ sr$^{-1}$. 

The fluctuation amplitude, $\delta$$I_{\nu}$, of the clustering component 
relative to the mean CIB level, $<I_{\nu}>$, can be estimated from the power 
spectrum at a given wavenumber to be \citep{kashlinsky05};
\begin{equation}
\delta I_{\nu}=\sqrt(k^{2}P_{2}(k)/2\pi)
\end{equation}
Using this relation, our measured fluctuation power 
of the galaxy clustering $P_{c}$ corresponds to a relative fluctuation 
amplitude of $\delta I_{\nu}$/$<I_{\nu}>$$=(2.3\pm1.7)\times10^{-3}$. 

It is notable that the CIB distribution at 90 $\micron$ after removing the point 
sources to a faint flux level is unexpectedly smooth. It is also noteworthy 
that the AKARI data are of high enough quality to detect such very low 
fluctuation levels.

\subsubsection{SED of the CIB fluctuations}
\begin{figure}[!ht]
\begin{center}
   \resizebox{1.0\hsize}{!}{
     \includegraphics*[0,0][900,600]{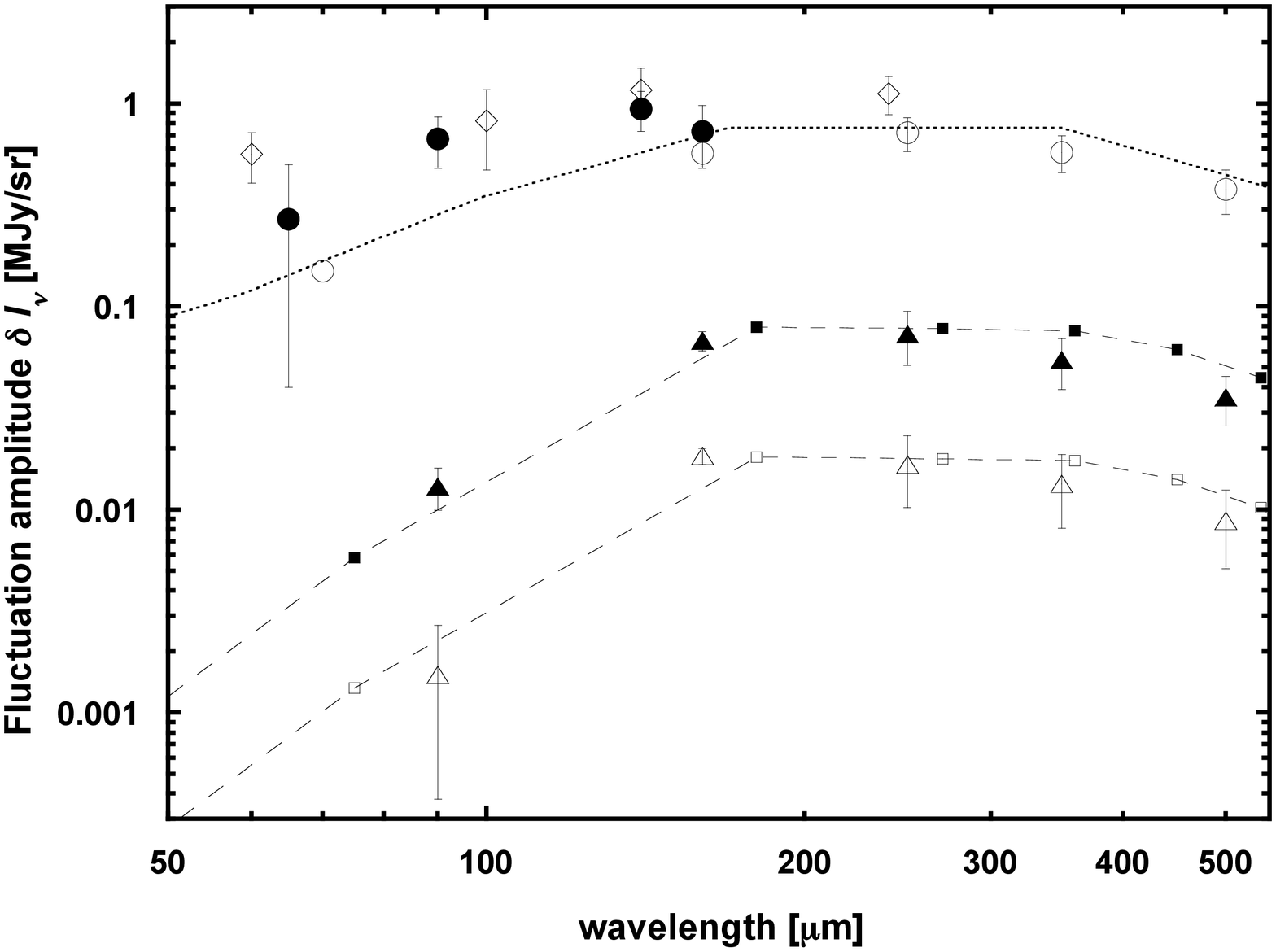}
   }
\caption{
The spectral energy distribution (SED) of the CIB fluctuations measured 
with AKARI at 90 $\micron$, Spitzer at 160 $\micron$ \citep{lagache07} and 
BLAST at 250, 350 and 500 $\micron$ \citep{viero09} for the shot noise levels 
at $k=0.5$ arcmin$^{-1}$ (filled triangles) and the galaxy clustering at 
$k=0.03$ arcmin$^{-1}$ (open triangles). The fluctuation power is converted 
to linear brightness scale as the fluctuation amplitude, 
$\delta$$I(k)=\sqrt(k^{2}P_{2}(k)/2\pi)$. 
The fluctuations are compared with the absolute CIB levels, $I_{\nu}$, 
same as in Figure~\ref{fig:cirb}; AKARI (filled circles), COBE/DIRBE 
(diamonds), Spitzer and BLAST (open circles). The model CIB by 
\citet{lagache04} is also indicated (dotted line) as a reference SED of 
the CIB composed of the 24 $\micron$ sources, which are stacked with 
the Spitzer and BLAST data to obtain the mean CIB levels.
The dashed lines with data points of the filled squares and the open 
squares are SED of Arp220 from \citet{klaas01}, which is redshifted to be 
at $z=2$ and vertically scaled to fit the CIB fluctuations data for the shot 
noise (upper) and the galaxy clustering (lower).
}\label{fig:cirb_fluct}
\end{center}
\end{figure}

A study of the  SED of the CIB fluctuations should give constraints on the redshift 
distribution and SED of the underlying galaxies and provide important 
information on their origin. Figure~\ref{fig:cirb_fluct} summarizes 
the CIB fluctuations measured with AKARI, Spitzer \citep{lagache07} and 
BLAST \citep{viero09}, for shot noise levels at $k=0.5$ arcmin$^{-1}$ 
and for galaxy clustering at $k=0.03$ arcmin$^{-1}$, with the fluctuation 
amplitude $\delta I_{\nu}$. For comparison, the measured mean CIB levels 
and the model CIB by \citep{lagache04} as in Figure~\ref{fig:cirb_fluct} 
are also shown. 

If the clustering components measured by AKARI, Spitzer and BLAST 
originate from the same underlying galaxies, the combined spectrum should 
be comparable to the SED of the most dominant galaxy population. 
In Figure~\ref{fig:cirb_fluct} we compar the combined spectrum of the CIB 
fluctuations with the SED of Arp220 taken from \citep{klaas01}, which is 
redshifted to $z=2$ and scaled to fit the data. Although there is a non-negligible 
discrepancy between the two spectra, the overall colors are very similar. 
Therefore, the clustering component is likely to be formed by such luminous 
infrared galaxies at redshifts of $z>1$, which are all resolved at 24 $\micron$ 
and account for $\sim100\%$ of the CIB in the submillimeter 
\citep{devlin09, pascale09, viero09}. This result is consistent in that 
the color of the clustering component is very similar to the color of the CIB 
in the Spitzer 160 $\micron$ band and the BLAST bands and  also consistent 
with the mean CIB levels in those bands which agree well with the model CIB 
for the 24$\micron$-selected galaxies. 

It is noteworthy that the shot noise as shown in Figure~\ref{fig:cirb_fluct} 
has a very similar spectrum to that of the clustering and to Arp220, in spite 
of the fact that the shot noise depends strongly on the limiting flux of the point 
source removal, which are quite different among the missions. This result 
implies that the brightest underlying galaxies in the field are tracing the clustering 
structure, because the CIB fluctuation is more sensitive to the brighter end of 
the source counts than the mean CIB level, as discussed in the previous section. 

Another remarkable result on the SED of the CIB fluctuations is that 
the relative fluctuation amplitude at 90 $\micron$, $\delta I_{90}$/$<I_{90}>$, 
is quite small compared with those at the other wavelengths, as summarized 
in Table~\ref{tbl-5}. This is caused by both very red color of the CIB 
fluctuations and the high CIB level at 90 $\micron$ exceeding the integrated 
flux of the 24$\micron$-selected galaxies, which can account for the mean 
CIB levels at longer wavelengths. According to this result, the CIB should 
have a spectrum peaking around 90 $\micron$if it is  not to contribute much at 
160 $\micron$ and not to be detected at 24 $\micron$. The underlying galaxies 
building the CIB should also be individually faint and numerous to keep 
the mean CIB levels high while keeping the CIB fluctuations low. 

Although hot high-$z$ sources with far-infrared luminosities of the order 
of $10^{13} L_{\sun}$ discussed in \citet{brain04} can considerably 
contribute to CIB at 90 $\micron$, such very luminous and less numerous 
galaxies would largely contribute to the CIB fluctuations rather than the mean 
CIB brightness. According to the measured CIB properties as discussed 
above, the bulk CIB may consist of hot high-$z$ sources of lower luminosity 
but higher surface density. A simple way of reducing the CIB fluctuations 
without the SED change is to assume higher redshifts and higher temperatures 
by keeping $T_d/(1+z)$ to be a constant, and this could be the case from our result. 

To obtain a definite conclusion to the origin of the CIB excess in the AKARI band, 
further investigation is required. To prove the hot high-$z$ source hypothesis, 
large aperture space telescope missions, such as SPICA \citep{nakagawa08}, 
are required to resolve the far-infrared CIB into point sources to very low flux 
levels (below $0.1$ mJy \citep{swinyard09}). Spectral measurement of 
the continuous SED of the CIB over a wide infrared range would also be a powerful 
discriminator of the galaxy population. Such observations can be simply done 
with a small aperture telescopes on small satellites or sub-orbital missions such 
as rocket experiments. Future space missions to obtain a direct measurement of 
the mid-infrared CIB mitigating the contamination by zodiacal emission require 
out-of-ecliptic missions or deep space missions beyond the asteroid belt 
(as the source of zodiacal dust), which will be powerful enough to probe into 
the underlying galaxies which compose of the CIB excess \citep{matsuura02}.

\section{Conclusions}
As the result of a cosmological survey in the ADF-S field, we have successfully 
measured the absolute sky brightness in the four photometric bands of AKARI 
at 65, 90, 140 and 160 $\micron$. AKARI's high sensitivity for diffuse 
emission, the field selection to minimize the foreground, and careful 
foreground estimates have enabled us to detect the CIB as an isotropic emission 
component. The high angular resolution of AKARI was also essential for 
removing the foreground galaxies as point sources.

The measured CIB brightness in all AKARI bands is consistent with 
the previous COBE/DIRBE results, and the 90 $\micron$ brightness is about 
twice higher than a simple interpolation from the Spitzer results at 70 and 
160 $\micron$ obtained by stacking analysis of 24 $\micron$-selected 
sources. The result suggests the exstence of new population(s) with high 
temperature spectra in addition to the previously explored infrared galaxy 
population.

By a power spectrum analysis of the 90 $\micron$ map after removing 
the point sources down to $\sim$20 mJy, we found the shot noise 
component due to the underlying galaxies. We also detected the signature 
of galaxy clustering at angular scales of a few tens $arcmin$. 
As a result of a comparison of our measured power spectrum with the Spitzer 
result at 160 $\micron$ and the BLAST result in the submillimeter range 
on the same angular scales, the SED of the clustering component is found to 
be as red as that of ULIRGs at high redshifts, and the main bulk of the CIB 
may be composed of another population with little contribution to 
the clustering power.

Our results provide new information on the evolution and the clustering 
properties of infrared galaxies. To obtain a definite conclusion on the origin 
of CIB, further investigation is required, e.g., the study of the SEDs of infrared 
galaxies in detail and a cross-correlation study with multi-band images. 
Resolving the excess brightness into point sources and measuring the CIB 
spectrum continuously over a wide range would be the main target of on-going 
and future infrared missions; Herschel, SPICA and other satellites, deep space 
missions, and sub-orbital programs.

\acknowledgements
We thank to L. Levenson for his help to calculate the zodiacal foreground 
in the far-infared based on the model in \citet{wright98}. This research is based 
on observations with AKARI, a JAXA project with the participation of ESA. 
This work was supported by KAKENHI (19540250 and 21111004).

\clearpage


\clearpage

\begin{table}
\begin{center}
\caption{Observation log.\label{tbl-1}}
\begin{tabular}{ccccc}
\tableline\tableline
Field &Center coord. (J2000) &Observation date &AOT(parameter) &$N_{obs}$\\
\tableline
ADF-S &$4^{h}44^{m}00^{s}$, $-53\degr20\arcmin00\arcsec$ &2006 Jul - 2006 Aug &FIS--02(2.0;15;1) &53\\
ADF-S &$4^{h}44^{m}00^{s}$, $-53\degr20\arcmin00\arcsec$ &2007 Jan - 2007 Feb &FIS--02(2.0;15;1) &63\\
ADF-S &$4^{h}44^{m}00^{s}$, $-53\degr20\arcmin00\arcsec$ &2007 Jul - 2007 Aug &FIS--01(1.0;15;1) &84\\
LH-PV\tablenotemark{a} &$10^{h}50^{m}36^{s}$, $57\degr27\arcmin09\arcsec$ &2006 May &FIS--02(2.0:15;1) &11\\
LH-MP\tablenotemark{b} &$10^{h}50^{m}25^{s}$, $57\degr36\arcmin37\arcsec$ &2007 May &FIS--01(2.0;08;1) &3\\
NEP-DT\tablenotemark{c} &$17^{h}55^{m}24^{s}$, $66\degr37\arcmin32\arcsec$ &2007 Oct &FIS--01(2.0;08;1) &4\\
\tableline
\end{tabular}
\tablenotetext{a}{Performance verification (PV) phase observations}
\tablenotetext{b}{Mission program (MP) observations}
\tablenotetext{c}{Director time (DT) observations}
\end{center}
\end{table}

\clearpage

\begin{table}
\begin{center}
\caption{Summary of foregrounds and mean CIB levels in ADF-S
\label{tbl-2}}
\begin{tabular}{cccccc}
\tableline\tableline
Wavelength &Total\tablenotemark{a} &Zodiacal\tablenotemark{b} 
&Cirrus\tablenotemark{c} &CIB\tablenotemark{d} &Models\tablenotemark{e}\\
($\micron$) &(MJy sr$^{-1}$) 
&(MJy sr$^{-1}$) &(MJy sr$^{-1}$) &(MJy sr$^{-1}$) &(MJy sr$^{-1}$)\\
\tableline
65 &4.31$\pm$0.04 &3.97 &0.09 &0.27$\pm$0.03$\pm$0.20 &0.1$-$0.2\\
90 &3.57$\pm$0.02 &2.72 &0.22 &0.67$\pm$0.05$\pm$0.14 &0.2$-$0.5\\
140 &2.36$\pm$0.16 &0.90 &0.57 &0.94$\pm$0.16$\pm$0.05 &0.5$-$1.4\\
160 &2.02$\pm$0.20 &0.77 &0.60 &0.73$\pm$0.21$\pm$0.04 &0.5$-$1.6\\
\tableline
\end{tabular}
\tablenotetext{a}{Mean sky brightness of all observations derived with 
the in-flight calibration factor. The error denotes statistical error only.}
\tablenotetext{b}{Mean zodiacal brightness of all observations calculated with 
the zodiacal emission model \citep{kelsall98}.}
\tablenotetext{c}{Mean cirrus brightness in the entire field converted from 
SFD map at 100$\micron$ \citep{finkbeiner00} by using the linear correlation 
for the short wavelength bands and assuming the blackbody spectrum with 
$T$=18.2 K and $n$=2 for the long wavelength bands.}
\tablenotetext{d}{In the short wavelength bands, zero-intercept of linear correlation 
between AKARI and SFD maps to subtract cirrus brightness. In the long wave 
bands, the mean cirrus brightness was subtracted from the zodi-subtracted 
mean brightness. The gain factor measured in the laboratory is used. 
The first error includes both statistical error and calibration error, 
and the second error is uncertainty of the zodi-model.}
\tablenotetext{e}{CIB brightness range predicted from galaxy evolution models, 
\citet{lagache04, pearson07, franceschini08, takeuchi10}.}
\end{center}
\end{table}

\clearpage

\begin{table}
\begin{center}
\caption{Field variance of CIB brightness \label{tbl-3}}
\begin{tabular}{cccccc}
\tableline\tableline
Field &$N_{HI}$\tablenotemark{a} &$I_{65}$\tablenotemark{b} 
&residual\tablenotemark{c} &$I_{90}$\tablenotemark{b} &residual\tablenotemark{c}\\
 &($10^{20}$ cm$^{-2}$) &(MJy sr$^{-1}$) & &(MJy sr$^{-1}$) & \\
\tableline
ADF-S &0.67 &0.329$\pm$0.003 &0.21$\pm$0.02 &0.835$\pm$0.004 &0.54$\pm$0.04\\
LH-MP &0.54 &0.311$\pm$0.057 &0.22$\pm$0.06 &0.873$\pm$0.021 &0.64$\pm$0.04\\
NEP-DT &3.73 &0.986$\pm$0.023 &0.33$\pm$0.09 &2.038$\pm$0.058 &0.42$\pm$0.22\\
average & & &0.22$\pm$0.02 & &0.59$\pm$0.03\\
\tableline
\end{tabular}
\tablenotetext{a}{Taken from the LAB HI 20 cm survey \citep{kalberla05}.}
\tablenotetext{b}{The mean brightness after subtraction of the zodiacal foreground 
with the errors estimated from the variance for a number of observations in each field.}
\tablenotetext{c}{Obtained by subtraction of the HI-correlated component, 
assuming the conversion factor at 100 $\micron$ to be $0.6\pm0.1$ 
MJy sr$^{-1}/10^{20}$cm$^{-2}$ and extrapolating it to each wavelength 
by the blackbody spectrum with $T$=18.2 K and $n$=2.}
\end{center}
\end{table}

\clearpage

\begin{table}
\begin{center}
\caption{Small-scale fluctuations of foregrounds and the CIB \tablenotemark{a} 
\label{tbl-4}}
\begin{tabular}{ccccc}
\tableline\tableline
Zodiacal\tablenotemark{b} &Cirrus\tablenotemark{c}
&CIB $P_{shot}$\tablenotemark{d} &Models\tablenotemark{e} &ISO\tablenotemark{f}\\
\tableline
$<0.6\times10^{2}$ &$<0.1\times10^{2}$ &$3.6\pm0.2\times10^{2}$ 
&$(1.1-7.2)\times10^{3}$ &$2\times10^{4}$\\
\tableline
\end{tabular}
\tablenotetext{a}{Power pectral density at 90 $\micron$ in Jy$^{2}$ sr$^{-1}$.}
\tablenotetext{b}{$<1\%$ of the mean brightness at angular scale of 1 arcmin 
\citep{abraham98}.}
\tablenotetext{c}{Cirrus fluctuation power at $k>0.2$ arcmin$^{-1}$ 
estimated from the measured $P_{0}$.}
\tablenotetext{d}{The mean shot noise level in $k=0.2-0.8$ arcmin$^{-1}$ 
after point-source removal.}
\tablenotetext{e}{Galaxy evolution models, \citet{lagache04, jeong06, 
pearson07, franceschini08, takeuchi10}.}
\tablenotetext{f}{ISO result at 90 $\micron$ after removing galaxies brighter 
than 30 mJy, \citet{matsuhara00}.}
\end{center}
\end{table}

\clearpage

\begin{table}
\begin{center}
\caption{Measured SED of the clustering component of the CIB fluctuations \label{tbl-5}}
\begin{tabular}{ccccc}
\tableline\tableline
Wavelength ($\micron$) &$\delta I_{\nu}$ (MJy sr$^{-1}$) &$\delta I_{\nu}$/$<I_{\nu}>$ &Reference\\
\tableline
  90 &(1.5$\pm$1.1)$\times$10$^{-3}$ &(2.3$\pm$1.7)$\times$10$^{-3}$ &This work\\
160 &(1.8$\pm$0.2)$\times$10$^{-2}$ &(3.2$\pm$0.3)$\times$10$^{-2}$ &1, 2\\
250 &(1.7$\pm$0.3)$\times$10$^{-2}$ &(2.3$\pm$0.4)$\times$10$^{-2}$ &3, 4\\
350 &(1.3$\pm$0.3)$\times$10$^{-2}$ &(2.3$\pm$0.5)$\times$10$^{-2}$ &3, 4\\
500 &(0.9$\pm$0.2)$\times$10$^{-2}$ &(2.3$\pm$0.6)$\times$10$^{-2}$ &3, 4\\
\tableline
\end{tabular}
\tablenotetext{}{\textbf{Notes.} The fluctuation amplitude of the clustering component 
estimated from Eq. 6 for $k=0.03$ arcmin$^{-1}$. }
\tablenotetext{}{\textbf{References.} (1)\citet{dole06}; (2)\citet{lagache07}; 
(3)\citet{marsden09}; (4)\citet{viero09}}
\end{center}
\end{table}

\end{document}